\documentclass[aps,preprint,tightenlines,floatfix,superscriptaddress,nofootinbib,showpacs]{revtex4-1}
\pdfoutput=1
\usepackage{graphicx}
\usepackage{amsfonts}
\usepackage{hhline}
\usepackage{multirow}
\usepackage{array}
\usepackage{amsmath}
\usepackage{amssymb}
\usepackage{float}
\usepackage[lofdepth,lotdepth]{subfig}
\usepackage[countmax]{subfloat}
\usepackage{cancel}
\def\beq{\begin{equation}}
\def\eeq{\end{equation}}
\def\bea{\begin{eqnarray}}
\def\eea{\end{eqnarray}}

\def\roughly#1{\mathrel{\raise.3ex\hbox
{$#1$\kern-.75em\lower1ex\hbox{$\sim$}}}}

\def\tauprocess{\tau\rightarrow K\pi\pi\nu_{\tau}}
\def\tauminusprocess{\tau^-\rightarrow K^-\pi^-\pi^+\nu_{\tau}}

\usepackage[normalem]{ulem}
\usepackage{color}
\definecolor{BrickRed}{cmyk}{0,0.89,0.94,0.28}
\definecolor{DarkGreen}{cmyk}{1,0,1,0.5}
\definecolor{Blue}{cmyk}{1,1,0,0}
\definecolor{BurntOrange}{cmyk}{0,0.51,1,0}

\def\soutdl{\bgroup\markoverwith{\textcolor{BrickRed}{\rule[0.5ex]{2pt}{0.4pt}}}\ULon}
\def\soutps{\bgroup\markoverwith{\textcolor{DarkGreen}{\rule[0.5ex]{2pt}{0.4pt}}}\ULon}
\def\soutkk{\bgroup\markoverwith{\textcolor{Blue}{\rule[0.5ex]{2pt}{0.4pt}}}\ULon}
\def\soutas{\bgroup\markoverwith{\textcolor{BurntOrange}{\rule[0.5ex]{2pt}{0.4pt}}}\ULon}

\begin{document}

\vspace*{2cm}

\title{\boldmath Probing sensitivity to charged scalars through partial differential widths: $\tau\rightarrow K\pi\pi\nu_{\tau}$ decays}

\def\tayloru{\affiliation{\it Physics and Engineering Department,
    Taylor University, \\ 236 West Reade Ave., Upland, IN 46989, USA \vspace*{8mm}}}
\def\laplata{\affiliation{\it IFLP, CONICET -- Dpto. de F\'{\i}sica,
    Universidad Nacional de La Plata, C.C. 67, 1900 La Plata,
    Argentina \vspace*{1mm}}}

\author{Nicolas Mileo \vspace*{4mm}}
\email{mileo@fisica.unlp.edu.ar}
\laplata

\author{Ken Kiers}
\email{knkiers@taylor.edu}
\tayloru

\author{Alejandro Szynkman}
\email{szynkman@fisica.unlp.edu.ar}
\laplata

\date{\today \\\bigskip\bigskip}

\begin{abstract}
\vspace*{4mm} We define and test $\mathrm{CP}$-even and
$\mathrm{CP}$-odd partial differential widths for the process
$\tauprocess$ assuming that an intermediate heavy
charged scalar contributes to the decay amplitude. Adopting a model-independent approach, we use a Monte Carlo simulation in order to study the number of events needed to recover information on the new
physics from these observables. Our analysis of the $\mathrm{CP}$-odd
observables indicates that the magnitude of $f_H \eta_P$, which is related to the
new-physics contribution, can be recovered with an uncertainty smaller
than $3\%$ for $3\times 10^6$ events. This number of events would also allow one to retrieve certain parameters appearing in the SM amplitude at the percent level. In addition, we discuss the possibility of using the
proposed observables to study specific models involving two Higgs doublets, such as the aligned two-Higgs-doublet model (A2HDM). This analysis is undertaken within the context of the upcoming Super B-factories, which are expected to provide a considerably larger number of events than that which was supplied  by the B-factories. Moreover, a similar set of observables could be employed to study other 
decay modes such as $\tau\rightarrow\pi\pi\pi\nu_{\tau},\,\tau\rightarrow KK\pi\nu_{\tau}$ and
$\tau\rightarrow KKK\nu_{\tau}$.
\end{abstract}

\maketitle

\section{Introduction}
With the discovery of a new boson $H$ by the ATLAS \cite{atlas} and CMS \cite{cms} collaborations, it is now very important to characterize this new particle in order to study the extent to which its features are in agreement with those predicted for the Higgs scalar within the Standard Model (SM). In particular, the spin of this new boson and its couplings to other particles have been carefully analyzed giving rise, with a high degree of confidence, to the conclusion that it has spin zero and that its couplings to the other particles are linearly correlated with their masses (see Refs.~\cite{Ellis1,Ellis2} and references therein). On the other hand, the possibility of an enlarged scalar spectrum is also being tested. In particular, from the high energy point of view, many searches for charged Higgs bosons decaying via $H\rightarrow \tau \nu_{\tau}$ have been performed by ATLAS and CMS (see, for example, Refs.~\cite{AtlasHtaunu}\nocite{CMSHtaunu}-\cite{CDF}). These searches have found the data to be consistent with the expected SM background and have set limits on the branching ratio of top quark decays to a $b$ quark and a charged Higgs boson. The effects of the presence of a charged Higgs boson can also be studied indirectly by means of low energy observables defined, for example, for leptonic and semileptonic decays involving $B,D^*,D,D_s,K$ and $\pi$ mesons \cite{PDG}. Such decays have been widely studied at the B-factories by the Belle and BaBar collaborations. Moreover, the fact that no new particle has been observed at the present time may suggest that the new physics (NP) scale is out of reach for the LHC. Indirect searches for physics beyond the SM become particularly important within this context.\par 

Among the various processes that can receive contributions from a
charged Higgs boson, the $\tau$ lepton decays can be
used to derive constraints on the scalar and pseudoscalar couplings of
a charged scalar to fermions. The fact that
$\mathrm{CP}$-violating effects are expected to be
  negligible within the SM means that a study of $\mathrm{CP}$-odd observables
  could reveal the presence of contributions from a charged Higgs
  boson, should the charged Higgs-fermion couplings violate
  $\mathrm{CP}$. Such an analysis has been carried out for the decay $\tau\rightarrow
K\pi\pi\nu_{\tau}$ in Ref.~\cite{KA}, where the presence of a charged
scalar contributing to the corresponding amplitude is assumed and two
types of $\mathrm{CP}$-asymmetries are defined in addition to the
usual partial rate asymmetry. In the present work, which extends the
  analysis of Ref.~\cite{KA}, we focus on the same decay $\tau\rightarrow K\pi\pi\nu_{\tau}$,
with the main goal being to define and test various $\mathrm{CP}$-even
and $\mathrm{CP}$-odd observables, on the one hand, and to study their
sensitivity to a NP contribution due to the presence of
a charged scalar, on the other. The decay under consideration, $\tau\rightarrow
K\pi\pi\nu_{\tau}$, only involves a pseudoscalar coupling of a charged
scalar to the up and strange quarks, in contrast to $\tau\rightarrow
K\pi\nu_{\tau}$, for instance, which exclusively probes the scalar
coupling \cite{CLEO}. It is also worth noting that the simplest $\tau$
decay with $\Delta S=1$ that probes the contribution arising from the
exchange of a charged scalar is $\tau\rightarrow K\nu_{\tau}$. In
fact, this decay involves exactly the same pseudoscalar coupling as
$\tau\rightarrow K\pi\pi\nu_{\tau}$, and then imposes constraints on
it.\par

For the analysis of the observables introduced below we use a large
number of Monte Carlo simulated events. The size of the 
Monte Carlo sample has been chosen within the context of the
upcoming Super B-factories, which are expected to significantly increase the luminosity as compared to
the B-factories. The aim of this analysis is to provide insight
into the number of events needed to extract information about the NP contribution as well as about the
SM contributions, including the anomalous Wess-Zumino (WZ) term.\par

Although our primary focus in the present work is on a
  model-independent treatment of charged-scalar contributions to
  $\tau\rightarrow K\pi\pi\nu_{\tau}$, it is useful also to consider a
  specific scenario. Many models include one additional Higgs doublet, so that
  a charged Higgs is present. In particular, in the so-called aligned
two-Higgs-doublet model (A2HDM), an alignment in flavour space of the Yukawa couplings of the two scalar
doublets is enforced, leading to the elimination of flavour-changing neutral
currents at tree level. This restrictive choice results in a highly predictive
phenomenology for this model, which has been carefully explored
(see Refs.~\cite{Pich1,Pich2,Branco}). Of particular
  interest to us is not only the fact that the A2HDM includes potential new
  sources of $\mathrm{CP}$ violation but also that it imposes very restrictive constraints due to the three-family universality of the proportionality constants arising from the alignment in flavour space. The partial differential
widths studied in this work can be considered as additional
observables to test the A2HDM, specifically within the context of the Super B-factories, in which the
possibility of extracting these distributions from the data is more plausible. In
this paper we briefly discuss the usefulness of the proposed
observables to probe the A2HDM.\par 

The remainder of this paper is organized as follows. In
Sec.~\ref{sec1} we write down the expression for the differential
width for the decay $\tau^-\rightarrow K^-\pi^-\pi^+\nu_{\tau}$ in
terms of the corresponding form factors, including both the NP and SM
contributions. By integrating the differential width weighted by
various angular functions, we define partial differential widths in
Sec.~\ref{sec2}. Section~\ref{sec3} introduces a set
of $\mathrm{CP}$-even and $\mathrm{CP}$-odd observables derived from
the weighted partial widths. The parameterization for the form
factors, along with the set of reference values used
later for the event simulation, are summarized in
Sec.~\ref{sec4}. The analysis of the proposed $\mathrm{CP}$-even and
$\mathrm{CP}$-odd observables is included in Sec.~\ref{sec5}. Finally,
in Sec.~\ref{sec6} the decay is considered in the context of the A2HDM
and in Sec.~\ref{sec7} some possibilities of testing the different
assumptions used during the paper are briefly discussed. We summarize
the main conclusions in Sec.~\ref{conclusions}.
The Appendix contains some details relevant for the statistical analysis.

\section{Differential width for $\tau^-\rightarrow K^-\pi^-\pi^+\nu_{\tau}$}
\label{sec1}
\vspace*{3mm}
We start with the effective Hamiltonian that accounts for the decay $\tau^-\rightarrow K^-\pi^-\pi^+\nu_{\tau}$ within the SM
\vspace*{3mm}
\begin{equation}
\label{eq1}
\mathcal{H}^{\mathrm{SM}}_{\mathrm{eff}}=\frac{G_F}{\sqrt{2}}\sin\theta_c\,[\bar{\nu}_{\tau}\gamma_{\mu}(1-\gamma_5)\tau]\,[\bar{s}\gamma^{\mu}(1-\gamma_5)u]+\mathrm{h.c.},
\vspace*{3mm}
\end{equation}
where $G_F$ is the Fermi constant and $\theta_c$ the Cabibbo angle. Possible NP effects due to a new charged scalar boson contributing to the decay may be included by adding the following terms to the effective Hamiltonian,
\vspace*{3mm}
\begin{equation}
\label{eq2}
\mathcal{H}^{\mathrm{NP}}_{\mathrm{eff}}=\frac{G_F}{\sqrt{2}}\sin\theta_c[\eta_S\,\bar{\nu}_{\tau}(1+\gamma_5)\tau\,\bar{s}u\,+\,\eta_P\,\bar{\nu}_{\tau}(1+\gamma_5)\tau\,\bar{s}\gamma_5 u]+\mathrm{h.c.},
\vspace*{3mm}
\end{equation}
where $\eta_S$ and $\eta_P$ are the scalar and pseudoscalar couplings, respectively. The hadronic matrix element $J^{\mu}\equiv \langle K^-(p_1)\pi^{-}(p_2)\pi^+(p_3)|\bar{s}\gamma^{\mu}(1-\gamma_5)u|0\rangle$ can be conveniently parameterized in terms of four form factors as follows, 
\vspace*{3mm}
$$\hspace*{-4mm}
J^{\mu} = \left[F_1(Q^2,s_1,s_2)(p_1-p_3)_{\nu}+F_2(Q^2,s_1,s_2)(p_2-p_3)_{\nu}\right]\!T^{\mu\nu}$$
\begin{equation}
\label{eq3}
+iF_3(Q^2,s_1,s_2)\epsilon^{\mu\nu\rho\sigma}p_{1\nu}p_{2\rho}p_{3\sigma}+F_4(Q^2,s_1,s_2)Q^{\mu},
\hspace*{-1cm}
\vspace*{3mm}
\end{equation}
where $Q^{\mu}=(p_1+p_2+p_3)^{\mu}$, $T^{\mu\nu}=g^{\mu\nu}-Q^{\mu}Q^{\nu}/Q^2$, $s_1=(p_2+p_3)^2$ and $s_2=(p_1+p_3)^2$ and where we adopt the convention $\epsilon_{0123}=+1$, as in Refs.~\cite{KA,Kuhn}. The functions $F_1-F_4$ are the form factors that arise from the different possible decay chains. $F_1$ and $F_2$ appear due to the decay chains involving the $K_1(1270)$ and $K_1(1400)$ resonances, $F_3$ is the anomalous Wess-Zumino term and $F_4$ is the scalar form factor, which is generally assumed to be negligible for this decay since there is no  pseudoscalar resonance through which the decay can proceed \cite{Decker}. The axial vector form factors $F_1$ and $F_2$ give the dominant contributions, while the anomalous vector form factor $F_3$ represents a subdominant contribution, as shown by numerical estimates \cite{CLEO}. The NP contribution coming from a scalar boson can be incorporated into the amplitude through the shift $F_4 \rightarrow \tilde{F}_4 = F_4 + f_H\eta_P/m_{\tau}$ \cite{KA}, where the pseudoscalar form factor $f_H$ is defined as 
\vspace*{3mm}
\begin{equation}
\label{eq4}
f_H=\langle K^-(p_1)\pi^-(p_2)\pi^+(p_3)|\bar{s}\gamma_5u|0\rangle\,. 
\vspace*{3mm}
\end{equation}
The starting point for our analysis will be the
differential width for the decay obtained from Eq.~(25) in
Ref.~\cite{KA} after integrating over the angle
$\theta$. The angle $\theta$ is defined in the rest
  frame of the tau; it is the angle between the direction of the hadrons (``$\vec{Q}$'') in that frame 
  and the direction of the tau in the laboratory frame. Performing the integration, we obtain
$$\frac{d\Gamma}{dQ^2ds_1ds_2d\gamma d\cos\beta}=\frac{A(Q^2)}{4\pi}\Big\{\Big[\frac{2}{3}\langle K_1\rangle+\langle K_2\rangle+\frac{1}{3}\langle \overline{K}_1\rangle\big(3\cos^2\beta-1\big)/2\Big]\big(|B_1|^2+|B_2|^2\big)+$$
$$\hspace*{2.6cm}+\Big[\frac{2}{3}\langle K_1\rangle+\langle K_2\rangle-\frac{2}{3}\langle\overline{K}_1\rangle\big(3\cos^2\beta-1\big)/2\Big]|B_3|^2+\langle K_2\rangle|B_4|^2$$
$$\hspace*{3.6cm} -\frac{1}{2}\langle\overline{K}_1\rangle\sin^2\beta\cos 2\gamma\big(|B_1|^2-|B_2|^2\big)+\langle\overline{K}_1\rangle\sin^2\beta\sin 2\gamma\,\mathrm{Re}(B_1B^*_2)$$
$$\hspace*{2cm} +2\langle\overline{K}_3\rangle\sin\beta\sin\gamma\,\mathrm{Re}(B_1B^*_3)+2\langle\overline{K}_2\rangle\sin\beta\cos\gamma\,\mathrm{Re}(B_1B^*_4)$$
$$\hspace*{2cm}+2\langle\overline{K}_3\rangle\sin\beta\cos\gamma\,\mathrm{Re}(B_2B^*_3)-2\langle\overline{K}_2\rangle\sin\beta\sin\gamma\,\mathrm{Re}(B_2B^*_4)$$
$$\hspace*{1.2cm}+2\langle\overline{K}_3\rangle\cos\beta\,\mathrm{Im}(B_1B^*_2)+\langle\overline{K}_1\rangle\sin 2\beta\cos\gamma\,\mathrm{Im}(B_1B^*_3)$$
\begin{equation}
\label{eq5}
\hspace*{1.4cm}-\langle\overline{K}_1\rangle\sin 2\beta\sin\gamma\,\mathrm{Im}(B_2B^*_3)+2\langle\overline{K}_2\rangle\cos\beta\,\mathrm{Im}(B_3B^*_4)\Big\} \, ,
\end{equation}
where
\vspace*{3mm}
\begin{equation}
\label{eq6}
A(Q^2)=\frac{G^2_F\sin^2\theta_c}{128(2\pi)^5}\frac{(m^2_{\tau}-Q^2)^2}{m^3_{\tau}Q^2},
\end{equation}
and
\begin{equation}
\label{eq7}
\langle K_i\rangle\equiv \frac{1}{2}\int^{\pi}_{0}K_i\sin\theta d\theta
\end{equation}
(and similarly for $\langle \overline{K}_i\rangle$); the
definitions of the $K_i$ and the $\overline{K}_i$ may be found
in Ref.~\cite{KA}. As described in Ref.~\cite{KA} (the definitions therein are identical to those in Ref.~\cite{Kuhn}), $\beta$ and $\gamma$ are Euler angles relating two coordinate systems used to specify the kinematics of the decay. Moreover, the functions $B_1-B_4$ are linearly related to the form factors as follows,\vspace*{3mm}
\begin{eqnarray}
\label{eq8}
B_1&=&[F_1(p_1-p_3)^x+F_2(p_2-p_3)^x]\\
\label{eq9}
B_2&=&(F_1-F_2)p^y_1\\
\label{eq10}
B_3&=&F_3\sqrt{Q^2}\,p^y_1p^x_3\\
\label{eq11}
B_4&=&\sqrt{Q^2}\left[F_4+\frac{f_H}{m_{\tau}}\eta_P\right].
\end{eqnarray}
Note that the form factors $F_i$ and $f_H$ are
  potential sources of strong phases, and that the only possible weak
  phase comes from the pseudoscalar coupling $\eta_P$. For future
reference, let us also define the quantity $\overline{B}_4$, which is
relevant for $\tau^+$ decays,
\begin{equation}
\label{eq12}
\overline{B}_4=\sqrt{Q^2}\left[F_4+\frac{f_H}{m_{\tau}}\eta^*_P\right].
\end{equation}
In fact, the differential width for the $\mathrm{CP}$-conjugate decay $\tau^+\rightarrow K^+\pi^+\pi^-\bar{\nu}_{\tau}$ can be obtained by replacing $B_4$ by $\overline{B}_4$ in Eq.~(\ref{eq5}) since the only source of CP violation appears in $B_4$ through the coupling $\eta_P$. For further details of the quantities involved within this section see Ref.~\cite{KA}.
\section{Weighted Differential Widths}
\label{sec2}
We now define observables that exploit the angular information 
that is available in the expression for the differential
width.  To do so, we employ weighting functions that allow us to 
isolate different contributions. Inspection of Eq.~(\ref{eq5}) reveals that it depends on nine different
functions of the angles $\beta$ and $\gamma$. These functions form an orthogonal
set; the functions, and their normalizations, are shown in Table
\ref{tabla1}.
\begin{table}[H]
\caption{Angular weighting factors. The $h_i(\gamma,\beta)$ functions form an othogonal set. The normalization factors are given in the third column.}
\centering
\begin{tabular}{c|cc}
\hline
\hline\\[-4mm]
$i$&$\quad h_i(\gamma,\beta)$&$\iint[h_i(\gamma,\beta)]^2\sin\beta d\gamma d\beta$\\[1mm]
\hline
$\,1\,$ & $\quad 1$ & $4\pi$\\
$\,2\,$ & $\quad 3\cos^2\beta-1$ & $16\pi/5$\\
$\,3\,$ & $\quad \sin^2\beta\cos 2\gamma$ & $16\pi/15$\\
$\,4\,$ & $\quad \sin^2\beta\sin 2\gamma$ & $16\pi/15$\\
$\,5\,$ & $\quad \sin\beta\sin\gamma$ & $4\pi/3$\\
$\,6\,$ & $\quad \sin\beta\cos\gamma$ & $4\pi/3$\\
$\,7\,$ & $\quad \cos\beta$ & $4\pi/3$\\
$\,8\,$ & $\quad \sin 2\beta\cos\gamma$ & $16\pi/15$\\
$\,9\,$ & $\quad \sin 2\beta\sin\gamma$ & $16\pi/15$\\
\hline
\hline
\end{tabular}
\label{tabla1}
\end{table}
The orthogonality of the functions means that different terms in Eq.~(\ref{eq5}) can be easily isolated by performing angular integrations of the differential width weighted by these angular functions. Hence, we can define nine weighted differential widths,  
\begin{equation}
\label{eq13}
\frac{d\Gamma_i}{dQ^2ds_1ds_2}\equiv\int\frac{d\Gamma}{dQ^2ds_1ds_2d\gamma\,d\cos\beta}h_i(\gamma,\beta)\sin\beta\,d\beta\,d\gamma,\quad\,i=1,...,9.
\end{equation} 
It is straightforward to perform the integrations in Eq.~(\ref{eq13}) using the information from Table \ref{tabla1}. The results for the various weighted differential widths are shown in Table \ref{tabla2}. The only weighted differential widths that include NP contributions are those with $i=1,5,6$ and $7$. Therefore, the remaining observables are clearly $\mathrm{CP}$-even.   
\begin{table}[H]
\caption{Weighted partial widths for the $\tau^-$ decay. The related expressions for the $\mathrm{CP}$-conjugate decay may be obtained by replacing $B_4$ by $\overline{B}_4$ everywhere it appears.}
\centering
\begin{tabular}{c|c}
\hline
\hline\\[-4mm]
$i$&$(d\Gamma_i/dQ^2ds_1ds_2)/A(Q^2)$\\[1mm]
\hline\\[-4mm]
1 & $\left(\frac{2}{3}\langle K_1\rangle+\langle K_2\rangle\right)\left(|B_1|^2+|B_2|^2+|B_3|^2\right)+\langle K_2\rangle|B_4|^2$\\[2mm]
2 & $\frac{2}{15}\langle\overline{K}_1\rangle\left(|B_1|^2+|B_2|^2-2|B_3|^2\right)$\\[2mm]
3 & $-\frac{2}{15}\langle\overline{K}_1\rangle\left(|B_1|^2-|B_2|^2\right)$\\[2mm]
4 & $\frac{4}{15}\langle\overline{K}_1\rangle\,\mathrm{Re}(B_1B^*_2)$\\[2mm]
5 & $\frac{2}{3}\langle\overline{K}_3\rangle\,\mathrm{Re}(B_1B^*_3)-\frac{2}{3}\langle\overline{K}_2\rangle\,\mathrm{Re}(B_2B^*_4)$ \\[2mm]
6 & $\frac{2}{3}\langle\overline{K}_3\rangle\,\mathrm{Re}(B_2B^*_3)+\frac{2}{3}\langle\overline{K}_2\rangle\,\mathrm{Re}(B_1B^*_4)$\\[2mm]
7 & $\frac{2}{3}\langle\overline{K}_3\rangle\,\mathrm{Im}(B_1B^*_2)+\frac{2}{3}\langle\overline{K}_2\rangle\,\mathrm{Im}(B_3B^*_4)$\\[2mm]
8 & $\frac{4}{15}\langle\overline{K}_1\rangle\,\mathrm{Im}(B_1B^*_3)$\\[2mm]
9 & $-\frac{4}{15}\langle\overline{K}_1\rangle\,\mathrm{Im}(B_2B^*_3)$ \\[1mm]
\hline
\hline
\end{tabular}
\label{tabla2}
\end{table}
\section{Observables}
\label{sec3}
Since we are assuming that $\mathrm{CP}$ is violated via the pseudoscalar coupling, the $\tau^-$ and $\tau^+$ distributions are not expected to be identical. There are in principle two ways to proceed. The first is to analyze the observables in Table \ref{tabla2} twice, once for the $\tau^-$ decay and once for the $\tau^+$ decay. Another possibility is to perform an analysis separately for the \emph{sum} and the \emph{difference} of the distributions. We will follow the latter approach, since it has the advantage that the difference between the $\tau^-$ and $\tau^+$ distributions is sensitive to the presence of $\mathrm{CP}$ violation. We define then the following distributions 
\begin{equation}
\label{eq14}
\frac{d\Gamma^{\pm}_i}{dQ^2ds_1ds_2}\equiv \frac{1}{2}\left(\frac{d\Gamma_i}{dQ^2ds_1ds_2} \pm \frac{d\overline{\Gamma}_i}{dQ^2ds_1ds_2}\right),
\end{equation}
where $d\overline{\Gamma}_i/dQ^2ds_1ds_2$ is obtained from
$d\Gamma_i/dQ^2ds_1ds_2$ by the replacement $B_4\rightarrow \overline{B}_4$ 
(or, equivalently, $\eta_P\rightarrow
\eta^{*}_P$); see Eqs.~(\ref{eq11}) and (\ref{eq12}). We note that the
quantities $d\Gamma^{+}_i/dQ^2ds_1ds_2$ and
$d\Gamma^{-}_i/dQ^2ds_1ds_2$ are, by
construction, $\mathrm{CP}$-even and
$\mathrm{CP}$-odd, respectively. As was noted above, the only non vanishing
$\mathrm{CP}$-odd distributions are those with $i=1,5,6$ and $7$,
because the remaining weighted differential widths do not include NP contributions
(i.e., they are independent of $B_4$).\par
Let us first consider the distributions with $i=1$. After projection onto $Q^2, s_1$ or $s_2$, the
$\mathrm{CP}$-even distribution with $i=1$ gives
the $\mathrm{CP}$-average of the invariant mass distributions, which
  are the distributions that are usually studied in experimental
  analyses ~\cite{CLEO,Belle}. The corresponding expression is
obtained from Table \ref{tabla2},
\begin{equation}
\label{eq15}
\frac{d\Gamma^+_1}{dQ^2ds_1ds_2}=A(Q^2)\left(\frac{2}{3}\langle K_1\rangle+\langle K_2\rangle\right)\left(|B_1|^2+|B_2|^2+|B_3|^2\right)+\frac{\langle K_2\rangle}{2}(|B_4|^2+|\overline{B}_4|^2).
\end{equation} 
The $\mathrm{CP}$-odd distribution with $i=1$ is given by
\vspace*{2mm}
\begin{equation}
\vspace*{3mm}
\label{eq16}
\frac{d\Gamma^-_1}{dQ^2ds_1ds_2}=A(Q^2)\frac{\langle K_2\rangle}{2}(|B_4|^2-|\overline{B}_4|^2)=2A(Q^2)\langle K_2\rangle\frac{Q^ 2}{m_{\tau}}|F_4f_H\eta_P|\sin(\delta_4-\delta_H)\sin(\phi_H),
\end{equation}
where $\delta_4$ and $\delta_H$ denote the strong phases arising from
the SM scalar form factor $F_4$ and the pseudoscalar form factor
$f_H$, respectively, and $\phi_H$ is the weak phase
present in $\eta_P$. The above expression is related to the well known partial
rate asymmetry.  As was noted in Ref.~\cite{KA},
the partial rate asymmetry is
expected to be doubly suppressed due to the generally assumed
smallness of $F_4$ and $\eta_P$. Expressions for the remaining
non-zero $\mathrm{CP}$-even and $\mathrm{CP}$-odd weighted partial
differential widths may be found in Table \ref{tabla3}, where we have
made use of the following definitions,
\begin{eqnarray}
\label{eq17}
B^{(+)}_4 & \equiv & \frac{1}{2}(B_4+\overline{B}_4)=\sqrt{Q^2}\left[F_4+\frac{f_H}{m_{\tau}}\,\mathrm{Re}(\eta_P)\right]\\
\label{eq18}
B^{(-)}_4 & \equiv & \frac{1}{2}(B_4-\overline{B}_4)=\frac{\sqrt{Q^2}\,if_H}{m_{\tau}}\,\mathrm{Im}(\eta_P).
\end{eqnarray}
\vspace*{-2mm}
\begin{table}[H]
\caption{$\mathrm{CP}$-even (``$+$'') and $\mathrm{CP}$-odd (``$-$'')
  weighted partial widths. Several of the $\mathrm{CP}$-odd weighted
  partial widths are zero; these have been omitted.}
\centering
\begin{tabular}{c|c}
\hline
\hline\\[-2mm]
$i(\pm)$&$(d\Gamma^{\pm}_i/dQ^2ds_1ds_2)/A(Q^2)$\\[2mm]
\hline\\[-2mm]
$2(+)$ & $\frac{2}{15}\langle\overline{K}_1\rangle\left(|B_1|^2+|B_2|^2-2|B_3|^2\right)$\\[2mm]
$3(+)$ & $-\frac{2}{15}\langle\overline{K}_1\rangle\left(|B_1|^2-|B_2|^2\right)$\\[2mm]
$4(+)$ & $\frac{4}{15}\langle\overline{K}_1\rangle\,\mathrm{Re}(B_1B^*_2)$\\[2mm]
$5(+)$ & $\frac{2}{3}\langle\overline{K}_3\rangle\,\mathrm{Re}(B_1B^*_3)-\frac{2}{3}\langle\overline{K}_2\rangle\,\mathrm{Re}\!\left(B^*_2B^{(+)}_4\right)$ \\[2mm]
$6(+)$ & $\frac{2}{3}\langle\overline{K}_3\rangle\,\mathrm{Re}(B_2B^*_3)+\frac{2}{3}\langle\overline{K}_2\rangle\,\mathrm{Re}\!\left(B^*_1B^{(+)}_4\right)$\\[2mm]
$7(+)$ & $\frac{2}{3}\langle\overline{K}_3\rangle\,\mathrm{Im}(B_1B^*_2)-\frac{2}{3}\langle\overline{K}_2\rangle\,\mathrm{Im}\!\left(B^*_3B^{(+)}_4\right)$\\[2mm]
$8(+)$ & $\frac{4}{15}\langle\overline{K}_1\rangle\,\mathrm{Im}(B_1B^*_3)$\\[2mm]
$9(+)$ & $-\frac{4}{15}\langle\overline{K}_1\rangle\,\mathrm{Im}(B_2B^*_3)$ \\[2mm]
\hline
$5(-)$ & $-\frac{2}{3}\langle\overline{K}_2\rangle\,\mathrm{Re}\!\left(B^*_2B^{(-)}_4\right)$\\[2mm]
$6(-)$ & $\frac{2}{3}\langle\overline{K}_2\rangle\,\mathrm{Re}\!\left(B^*_1B^{(-)}_4\right)$\\[2mm]
$7(-)$ & $-\frac{2}{3}\langle\overline{K}_2\rangle\,\mathrm{Im}\!\left(B^*_3B^{(-)}_4\right)$\\[2mm]
\hline
\hline
\end{tabular}
\label{tabla3}
\end{table} 
Interestingly, from the definitions in Eqs.~(\ref{eq17}) and (\ref{eq18})
and the results in Table \ref{tabla3}, we note that
it does not seem to be possible to extract $F_4$ (by itself) from the
data when $\phi_H\neq \pm \pi/2$. In other words, there will always
be an admixture of $f_H\eta^R_P$,\footnote{From now on, we will use the superscripts $R$ and $I$ to denote the real and imaginary parts of a quantity, respectively.} and it will not be possible to distinguish them. However, if the coupling $\eta_P$ were purely imaginary, the
factor $B^{(+)}_4$ would only depend on the scalar form factor $F_4$
and then the $\mathrm{CP}$-even observables with $i=5,6$ and $7$ would
be useful for determining $F^{R,I}_4$.\par

In order to study the observables presented above (Table
\ref{tabla3}), we have made various assumptions that tend to simplify
the analysis, in a manner similar to the approach that was followed in
Ref.~\cite{KA}. First of all, note that the SM scalar form factor
$F_4$ is generally assumed to be small for $\tau\rightarrow
K\pi\pi\nu_{\tau}$, since there are no pseudoscalar resonances that
mediate this decay. Therefore, we will neglect this contribution by
setting $F_4=0$.  Furthermore, we will assume that $f_H$ has a flat
behaviour over the phase space (no $Q^2,s_1$ and $s_2$ dependence) and
does not contain strong phases. Thus, we set $f^I_H=0$. Under these
assumptions the $1(-)$ distribution is reduced to zero, as can be seen
from Eq.~(\ref{eq16}), while the $1(+)$ distribution becomes equal to the usual
(unweighted) differential width, as follows from Eqs.~(\ref{eq11}),
(\ref{eq12}) and (\ref{eq15}). Finally, in order to simplify and
separate the analysis of the $\mathrm{CP}$-even and $\mathrm{CP}$-odd
observables, we perform the analysis with $\phi_H=\pi/2$. For this
particular value, $B^{(+)}_4=0$ and the NP contribution is removed
from the $\mathrm{CP}$-even observables (see Eqs.~(\ref{eq15}) and
(\ref{eq17}), as well as Table \ref{tabla3}). To set an input value
for the quantity $|f_H\eta_P|$, we follow the approach
adopted in Ref.~\cite{KA}, where it is assumed that the NP
contribution to the width is hidden in the experimental uncertainty of
the branching ratio. As shown there, the experimental uncertainty is
saturated for $|f_H\eta_P|\simeq 17.9$. Thus, we take this value as a
reference input.  A few comments are in order at this point.
\begin{enumerate}
\item As is noted in Ref.~\cite{KA}, one way to
  obtain an estimate of the order of magnitude of $f_H$ is to compute $F_4$ within the context of Chiral
  Perturbation Theory (see Ref.~\cite{Decker}) and then to relate
  $f_H$ to $F_4$ via the quark equations of motion.  The latter step
  yields $f_H\sim Q^2 F_4/m_s$. A numerical study along
  these lines, with kinematical variables sampled appropriately over
  the relevant phase space, shows that $\langle |f_H|\rangle\sim 14$,
  with $76\%$ of the values falling within the range $7$-$21\,$. 
  Regarding the phase of $f_H$, one finds
  $\langle \arg\left(f_H\right)\rangle\simeq
  0.97\,\pi$, so that $\left|\langle\mathrm{Im}(f_H)\rangle\right|\ll
  \left|\langle\mathrm{Re}(f_H)\rangle\right|$. Thus, it appears to be
  reasonable to assume that $f_H$ is real.
\item The NP parameter $|\eta_P|$ should scale as
  $m_W^2/M^2$ due to the charged scalar propagator, with $m_W$ and $M$
  being the $W$ and charged scalar masses, respectively. If the
  charged scalar has electroweak couplings, it would be reasonable to
  assume that $\eta_P$ has a magnitude not exceeding unity.  
\item Combining the estimates from the above two comments, we obtain
  $|f_H\eta_P|\sim 14$, which is similar to our reference value
  $|f_H\eta_P|=17.9$.  As pointed
  out in Ref.~\cite{KA}, however, this estimate may well have large
  uncertainties due to the use of the quark equations of motion; a
  more realistic assumption would probably be to take $|f_H \eta_P|$
  to be in the range $1$-$10$.
\item The decay channel $\tau^-\rightarrow
  K^-\nu_{\tau}$ also involves the pseudoscalar coupling $\eta_P$, so
  that this process can in principle be used to constrain the NP contribution to
  $\tauprocess$.  It turns out, however, that the constraints derived
  from $\tau^-\rightarrow K^-\nu_{\tau}$ are very sensitive to the
  values used for the strange quark mass and its uncertainty. By
  performing a crude estimate that takes into account the
  uncertainties of the $K^-$ decay constant, $f_{K^-}$, and makes use
  of the quark equations of motion, we obtain the constraint
  $|\eta^I_P|<0.364$ (recall our assumption that $\phi_H=\pi/2$). We
  note that this bound was derived by using the value
  $m_s=0.095\,\mathrm{GeV}$. On the other hand, if the quark mass is
  replaced by the meson mass, one finds $|\eta^I_P|<1.878$. By
  combining these constraints with the assumption that $1<|f_H|<10$,
  we obtain two different bounds, namely $|f_H\eta^I_P|<3.64$ and
  $|f_H\eta^I_P|<18.78$. Therefore, the constraints provided by the
  decay channel $\tau^-\rightarrow K^-\nu_{\tau}$ are not conclusive
  enough to discard our input value.
\end{enumerate}
In much of the analysis that follows, we set
  $|f_H\eta_P|=17.9$.  With the above comments in mind, however, we
  also include some results for $f_H\eta_P=1.79\,e^{i\pi/4}$ in
  Sec.~\ref{sec5}.

\section{Parameterization of form factors}
\label{sec4}
We now introduce the parameterization of
the form factors $F_1-F_3$ appearing in the definitions of
the quantities $B_1-B_3$ in the expression for the differential width (see
  Eqs.~(\ref{eq5}) and (\ref{eq8})-(\ref{eq10})). We write the
form factors in terms of various Breit-Wigner functions in the
following manner,
\begin{eqnarray}
\label{eq21}
F_1(Q^2,s_1,s_2)&=&-\frac{2N}{3F_{\pi}}[C\cdot BW_{1270}(Q^2)+D\cdot BW_{1400}(Q^2)]\,BW_{K^*}(s_2)\\
\label{eq22}
F_2(Q^2,s_1,s_2)&=&-\frac{N}{\sqrt{3}F_{\pi}}[A\cdot BW_{1270}(Q^2)+B\cdot BW_{1400}(Q^2)]\,T^{(1)}_{\rho}(s_1)\\
\label{eq23}
F_3(Q^2,s_1,s_2)&=&\frac{N_3}{2\sqrt{2}\pi^2 F^3_{\pi}}BW_{K^*}(Q^2)\left[\frac{T^{(1)}_{\rho}(s_1)+\alpha\,BW_{K^*}(s_2)}{1+\alpha}\right].
\end{eqnarray}
The normalized Breit-Wigner propagators for the $K_1(1270)$ and the $K_1(1400)$ appearing in the axial vector form factors $F_1$ and $F_2$ are assumed to be \cite{CLEO},
\begin{equation}
\label{eq24}
BW_{K_1}(Q^2)=\frac{-m^2_{K_1}+im_{K_1}\Gamma_{K_1}}{Q^2-m^2_{K_1}+im_{K_1}\Gamma_{K_1}},
\end{equation}
where $m_{K_1}$ and $\Gamma_{K_1}$ denote the mass and width for the corresponding $K_1$ state. The Breit-Wigner propagators for the $K^*$ and $\rho$ are taken to have energy-dependent widths (see Refs.~\cite{CLEO,Decker2}),
\begin{equation}
\label{eq25}
BW_R(s)=\frac{-m^ 2_R}{s-m^ 2_R+i\sqrt{s}\Gamma_R(s)},
\end{equation}
with 
\begin{equation}
\label{eq26}
\Gamma_R(s)=\Gamma_R\frac{m^2_R}{s}\left(\frac{p}{p_R}\right)^3,
\end{equation}
where
\begin{eqnarray}
\label{eq27}
p&=&\frac{1}{2\sqrt{s}}\sqrt{[s-(m_1+m_2)^ 2][s-(m_1-m_2)^ 2]}\\
\label{eq28}
p_R&=&\frac{1}{2m_R}\sqrt{[m^ 2_R-(m_1+m_2)^ 2][m^ 2_R-(m_1-m_2)^ 2]}.
\end{eqnarray}
In the above expressions the decay of the resonance $R$ to two particles with masses $m_1$ and $m_2$ is assumed. For the $K^*$, a single resonance with an energy-dependent width is assumed while the expression for the $\rho$ includes two different resonances:
\begin{equation}
\label{eq29}
T^{(1)}_{\rho}(s_1)=\frac{BW_{\rho}(s_1)+\beta BW_{\rho\prime}(s_1)}{1+\beta}.
\end{equation}
To fix the reference values for the parameters $A-D$ in Eqs.~(\ref{eq21}) and (\ref{eq22}) we follow Ref.~\cite{CLEO}, where constraints  arising from the tabulated branching fractions of the $K_1$ resonances are imposed. Regarding the parameters $N$ and $N_3$ that regulate the contributions coming from the axial and anomalous form factors, respectively, we apply the criteria proposed in Ref.~\cite{KA}, in which $5\%$ of the $\tau\rightarrow K\pi\pi\nu_{\tau}$ width is ascribed to the $F_3$ term and the remaining $95\%$ to the $F_1$ and $F_2$ terms. For this computation, we have used the value of the branching ratio $\mathcal{B}(\tau\rightarrow K\pi\pi\nu_{\tau})$ obtained in \cite{Babar}, which is the most precise one at present (see Refs.~\cite{Belle,Belle2}). All the reference values related to the form factors $F_1-F_3$ used in our analysis are listed in Table \ref{tabla4}. Among them, those corresponding to the form factors $F_1$ and $F_2$ are based  on Ref.~\cite{CLEO}. We note that a more recent and precise value for the mass and the width of the $K_1(1270)$  resonance obtained in Ref.~\cite{Guler} from a signal-region fit for the channel $B^+\rightarrow J/\psi K^+ \pi^+ \pi^-$ is still in agreement with the input value used here. For the form factor $F_3$ we follow Ref.~\cite{Decker}, whereas for the $\rho$ and $\rho'$ resonances the input values are guided by Refs.~\cite{Finkemeier,Santamarina}.    

\begin{table}[H]
\caption{Input values for the parameters entering in the form factors $F_1-F_3$. The left table (a) lists the dimensionless parameters while the right table (b) shows the masses and widths of the various resonances, along with the pion decay constant ($F_{\pi}$).}
\centering
\subfloat[][]{
\centering
\begin{tabular}{c|c}
\hline
\hline\\[-3mm]
Parameter & Value\\[1mm]
\hline\\[-3mm]
$\alpha$ & $-0.2$\\[2mm]
$\beta$ & $-0.145$\\[2mm]
$A$ & $0.944$ \\[2mm]
$B$ & $0$\\[2mm]
$C$ & $0.195$\\[2mm]
$D$ & $0.266$\\[2mm]
$N$ & $1.4088$ \\[2mm]
$N_3$ & $1.4696$\\[1mm]
\hline
\hline
\end{tabular}}
\qquad
\subfloat[]{
\centering
\begin{tabular}{c|c}
\hline
\hline\\[-3mm]
Parameter & Value\\[1mm]
\hline\\[-3mm]
$F_{\pi}$ & $93.3\,\mathrm{MeV}$\\[2mm]
$m_{1270}$ & $1.254\,\mathrm{GeV}$\\[2mm]
$\Gamma_{1270}$ & $0.26\,\mathrm{GeV}$\\[2mm]
$m_{1400}$ & $1.463\,\mathrm{GeV}$\\[2mm]
$\Gamma_{1400}$ & $0.30\,\mathrm{GeV}$ \\[2mm]
$m_{K^*}$ & $0.892\,\mathrm{GeV}$\\[2mm]
$\Gamma_{K^*}$ & $0.050\,\mathrm{GeV}$\\[2mm]
$m_{\rho}$ & $0.773\,\mathrm{GeV}$\\[2mm]
$\Gamma_{\rho}$ & $0.145\,\mathrm{GeV}$ \\[2mm]
$m_{\rho\prime}$ & $1.370\,\mathrm{GeV}$\\[2mm]
$\Gamma_{\rho\prime}$ & $0.510\,\mathrm{GeV}$\\[1mm]
\hline
\hline
\end{tabular}}
\label{tabla4}
\end{table} 
\section{ANALYSIS}
\label{sec5}
In order to study the proposed observables we have performed two
different analyses. In the first we have tested the SM hypothesis.
In this case there are no $\mathrm{CP}$-violating effects present in this
decay and hence the $\mathrm{CP}$-odd observables in Table
\ref{tabla3} are zero. In the second analysis, we have performed various
fits of the distributions arising from all of the
observables in Table \ref{tabla3}. Both analyses have been implemented
by using our own Monte Carlo (MC) generator to
simulate several sets of events with different sizes. The main goal of
these two analyses is to estimate the number of events needed to
detect the presence of NP (in the case of the SM test) and to extract
the NP coupling (in the case of the fit to the $\mathrm{CP}$-odd
observables). Furthermore, the study of the
$\mathrm{CP}$-even observables aims to extract information about the
resonant structure of the decay and, in particular, of the anomalous
Wess-Zumino contribution.\par 

We have focused our analysis on a scenario in which the NP parameter is 
assumed to be hidden in the experimental uncertainty
of the branching ratio. Hence, as mentioned above, we have set
the input value for the NP contribution to be $17.9\,
e^{i\pi/2}$. In order to test the usefulness of the proposed observables when the NP
contribution is considerably reduced, we have also performed an
analysis of the $\mathrm{CP}$-odd observables in the case where
$f_H\eta_P = 1.79\,e^{i\pi/4}$. 
\subsection{Monte Carlo Simulation}
\label{sec5.1}
In order to simulate the distribution in Eq.~(\ref{eq5}), we have
constructed a Monte Carlo event generator by applying von
Neumann's acceptance-rejection technique. Once 
a set of events has been generated that is consistent with the
differential decay width, the different
observables can be obtained by using suitable estimators. By
employing our own event generator we are able to include
different contributions to the differential decay width and to choose their parameterization. Various sets of events have been generated 
for the decay $\tau^-\rightarrow
K^-\pi^+\pi^-\nu_{\tau}$ and for its
$\mathrm{CP}$-conjugate, $\tau^+\rightarrow
K^+\pi^+\pi^-\bar{\nu}_{\tau}$. The maximum number of events was taken
to be $3\times 10^6$ for the case in which the NP parameter $f_H\eta_P$
is equal to $17.9\,e^{i\pi/2}$ and
$10^6$ for the case with $f_H\eta_P=1.79\,e^{i\pi/4}$. Although the total number of
events in these simulations is beyond the scope of the B-factories, it
can be regarded as realistic within the context of the upcoming
Super B-factories, which are expected to increase the design
luminosity by approximately two orders of
magnitude. In fact, the design luminosity at SuperKEKB is $8\times
10^{35}\,\mathrm{cm}^{-2}\mathrm{s}^{-1}$ and an integrated luminosity
of $50\,\,\mathrm{ab}^{-1}$ is expected \cite{Akai}. Guided by the
analysis performed in Ref.~\cite{Belle} (which was based on data collected by the Belle detector at KEKB)
and taking into account the expected integrated luminosity at
SuperKEKB, we can estimate the expected number of
$\tauminusprocess$ events. A conservative estimate gives $\sim 5\times
10^6$, which is above the maximum number of events we have simulated
for the present analysis, $3\times 10^6$.\footnote{Even though the
  estimated number of events takes into account the possible
  backgrounds as well as the detector effects \cite{Akai}, these have
  not been considered during the present analysis.}\par
\begin{figure}[H]
\centering
\hspace*{-0.4cm}
\subfloat[][]{\includegraphics[width=0.52\textwidth,height=0.40\textwidth]{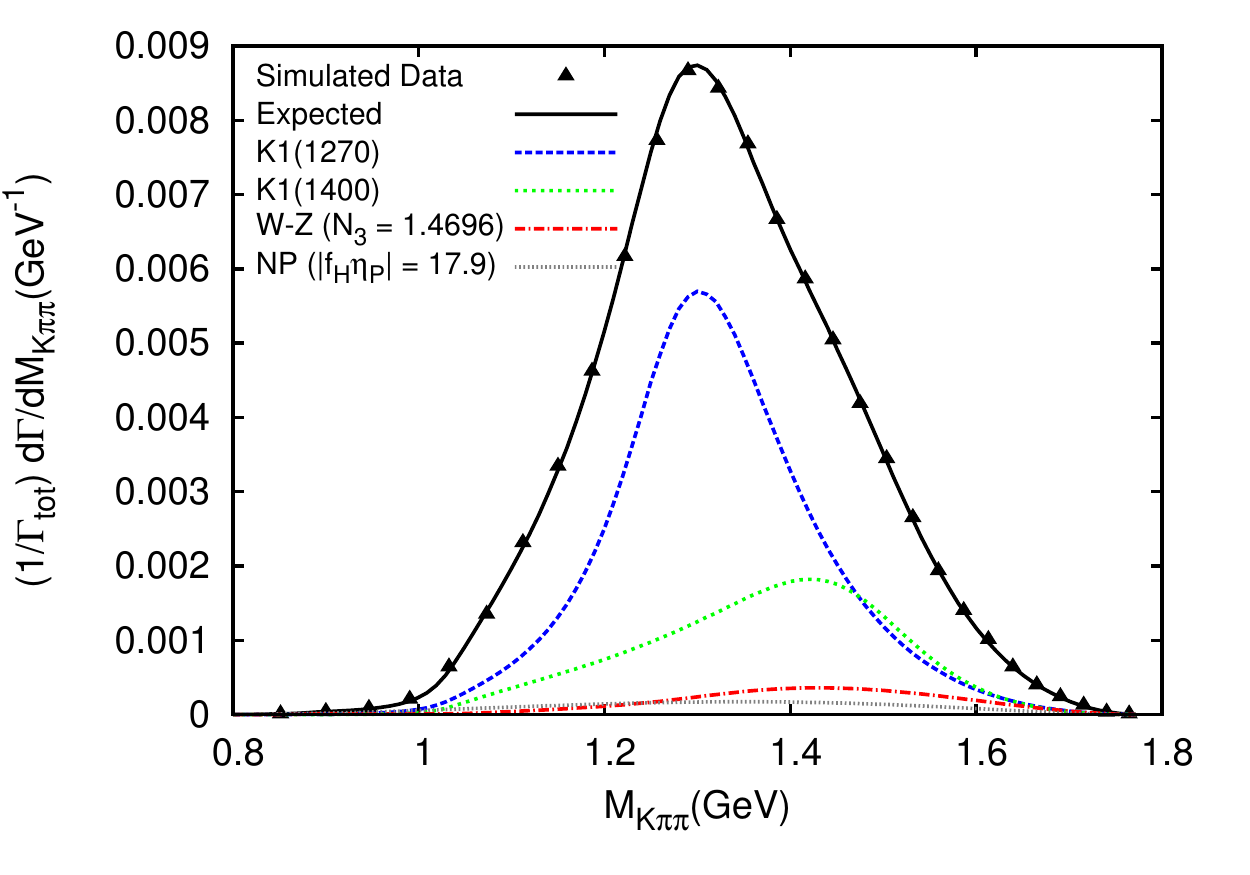}\label{fig1a}}
\subfloat[][]{\includegraphics[width=0.52\textwidth,height=0.40\textwidth]
{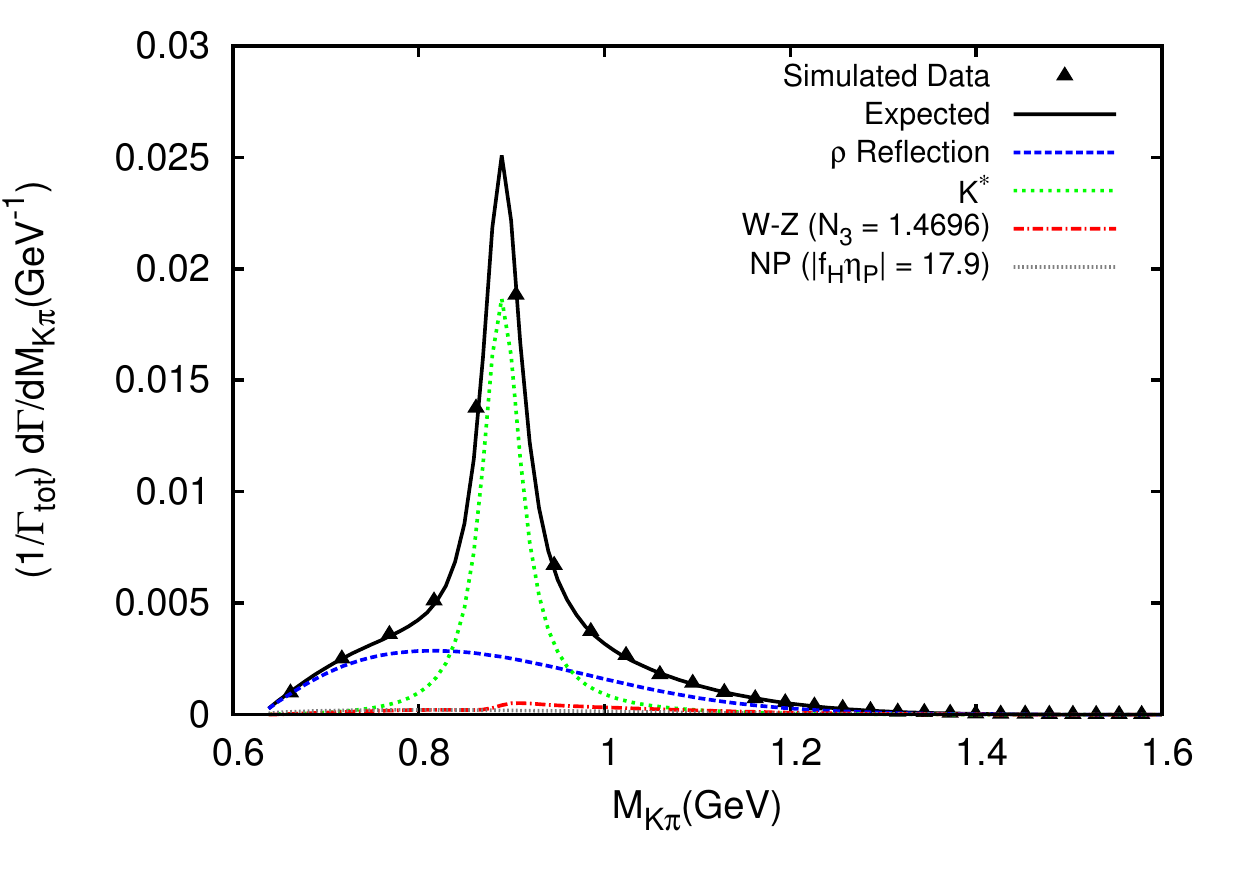} 
\label{fig1b}}
\\
\subfloat[][]{\centering \includegraphics[width=0.52\textwidth,height=0.40\textwidth]
{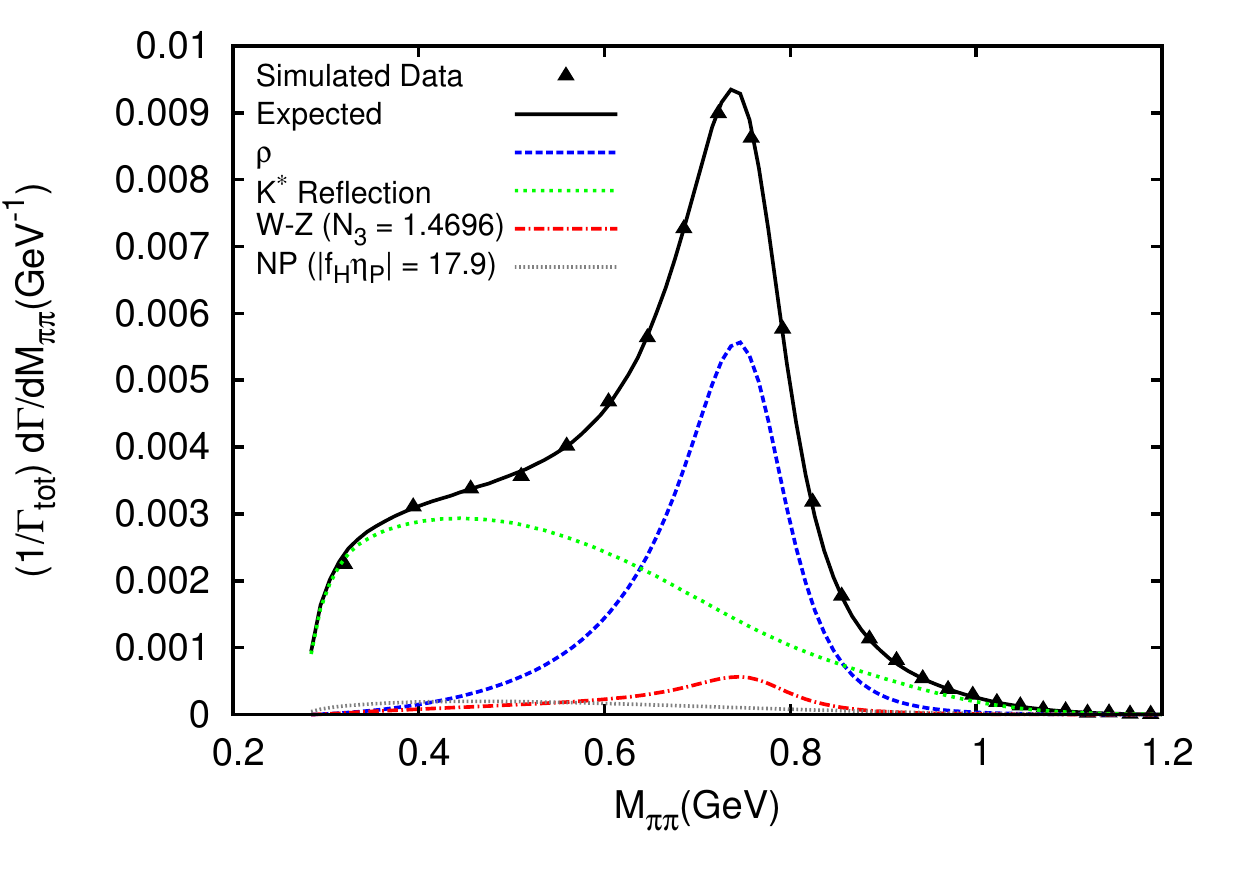}
\label{fig1c}}
\caption{Plots of the differential widths $d\Gamma/dM$, including the different contributions from the decay chains along with the simulated data points obtained by using our MC generator. The $|f_{H}\eta_P|$ curve displays the NP contribution.}
\label{fig1}
\end{figure}

 
As was noted in Sec.~\ref{sec3}, the pseudoscalar form factor has
been assumed to be real and the SM scalar contribution has been
neglected; thus, we have taken $f^I_H=F_4=0$ as inputs for the MC
simulation. The input values related to the form factors $F_1-F_3$ are
listed in Table \ref{tabla4}.
As a test of the consistency of our event generator, the usual
differential width distributions have been extracted from a set of
$1\times 10^ 5$ simulated events. As can be seen from Fig.~\ref{fig1},
the simulated distributions are in agreement with those 
obtained experimentally by the CLEO collaboration in Ref.~\cite{CLEO} and also with the
expected distributions based on numerical computations \cite{KA}. In
addition to the contributions involving the form factors $F_1$ and
$F_2$, the subdominant contribution from the W-Z term and the possible
NP contribution have been incorporated in the plots.
\subsection{SM Hypothesis test}
\label{sec5.2}
The fact that the $\mathrm{CP}$-odd observables $5(-),6(-)$ and $7(-)$
are zero if the NP contribution is absent (i.e., if $f_H\eta_P = 0$)
allows for a test of the SM hypothesis by performing a Pearson's
$\chi^2$-test.  To perform this test, we calculate
  $\chi^2$ for a particular observable $j(-)$ and then compute the
  quantity $P_j$, which is the probability that the hypothesis (the SM hypothesis in our case) would lead to a $\chi^ 2$ value greater than the one actually obtained, 
\begin{equation}
\label{eq1SMhyp}
\chi^2_j=\sum_{i=1}^{N_{\mathrm{bins}}}\left[\frac{d_x\Gamma^{-}_{j}(x_i)}{\sigma^{(j)}_i}\right]^2,\qquad P_j=\int_{\chi^2_j}^{\infty} f(z;n_d)dz,\qquad j=5,6,7.
\end{equation}
In the above expressions, $d_x\equiv
d/dx$ with $x\equiv Q^2,s_1,s_2$, $N_{\mathrm{bins}}$ is the number of
bins,\footnote{For the entire analysis we
  have used the conservative number of $20$ bins (see
  Ref.~\cite{Belle}).} $f(z;n_d)$ is the $\chi^2$ distribution for
$n_d$ degrees of freedom and $\sigma^{(j)}_i$ denotes the statistical
uncertainty in the $i$-th bin for the observable $j(-)$ (see
App.~\ref{sec8}). We remark that the values of the distributions in
the numerator of the expression for $\chi^2_j$ given in
Eq.~(\ref{eq1SMhyp}) are extracted from the simulations. It is worth
noting that this test is based on the
assumption that the SM contribution only includes strong phases and
therefore the only source of $\mathrm{CP}$-violation for the decay is
a weak phase present in the NP contribution. Hence, the test itself
does not depend on the particular value of the NP parameter, even when
its robustness actually does (as we will show later). Tables
\ref{tablapv1} and \ref{tablapv2} show results of the SM hypothesis
test performed using the observables $d\Gamma^{-}_{5,6,7}/dx$, with
different numbers of events, and taking $f_H\eta_P =
17.9\,e^{i\pi/2}$. 
\begin{table}[H]
\caption{$P$-values corresponding to the observables $5(-)$ and
  $7(-)$. The number of events considered is given in the first
  column.}
\label{tablapv1}
\begin{center}
\begin{tabular}{|c|c|c|c|c|c|c|}
\hhline{|=======|}\\[-4.5mm]
& \multicolumn{6}{|c|}{$P$-$\mathrm{values}$} \\
\hhline{|=======|}
$N_{\mathrm{ev}}/ 100,000$& $d\Gamma^{-}_5/dQ^2$&$d\Gamma^{-}_5/ds_1$&$d\Gamma^{-}_5/ds_2$&$d\Gamma^{-}_7/dQ^2$&
$d\Gamma^{-}_7/ds_1$&$d\Gamma^{-}_7/ds_2$\\[0.5mm]
\hhline{|=======|}
$5$ & $0.933$ & $0.754$ & $0.175$ & $0.0086$ & $0.168$ & $0.057$\\[0.5mm]
\hline
$10$ & $0.675$ & $0.361$ & $0.0018$ & $0.00015$ & $0.044$ & $0.00013$\\[0.5mm]
\hline
$15$ & $0.198$ & $0.062$ & $0.000015$ & $1.15\times 10^{-7}$ & $0.00033$ & $4.27\times 10^{-7}$\\[0.5mm]
\hline
$20$ & $0.286$ & $0.055$ & $2.73\times 10^{-7}$ & $2.78\times 10^{-10}$ & $8.14\times 10^{-6}$ & $9.76\times 10^{-11}$\\[0.5mm]
\hhline{|=======|}
\end{tabular}
\end{center}
\end{table}
\begin{table}[H]
\caption{$P$-values for the observable $6(-)$. The number of events is
  shown in the first
  column. Note that in
  this case fewer events were included in the simulations than were
used in the previous table.}
\label{tablapv2}
\begin{center}
\begin{tabular}{|c|c|c|c|}
\hhline{|====|}\\[-4.5mm]
& \multicolumn{3}{|c|}{$P$-$\mathrm{values}$} \\
\hhline{|====|}
$N_{\mathrm{ev}}/ 100,000$&
$d\Gamma^{-}_6/dQ^2$&$d\Gamma^{-}_6/ds_1$&$d\Gamma^{-}_6/ds_2$\\[0.5mm]
\hhline{|====|}
$1$ & $0.000024$ & $0.0076$ & $0.013$ \\[0.5mm]
\hline
$2$ & $1.27\times 10^{-14}$ & $1.05\times 10^{-7}$ & $1.25\times 10^{-6}$ \\[0.5mm]
\hline
$3$ & $< 10^{-17}$ & $6.66\times 10^{-16}$ & $3.55\times 10^{-15}$ \\[0.5mm]
\hline
$4$ & $< 10^{-17}$ & $< 10^{-17}$ &  $< 10^{-17}$ \\[0.5mm]
\hhline{|====|}
\end{tabular}
\end{center}
\end{table}
As shown in Table \ref{tablapv2}, the SM test for the observable $6(-)$ allows one 
to reject the SM hypothesis with as few as $10^5$
events. This is not the case for the other
$\mathrm{CP}$-odd observables, which are not useful
for rejecting the null hypothesis unless there are at
least $5\times 10^5$ events. In fact, one can use this $\chi^2$ test to rank
the various observables in terms of their sensitivity to the NP
contribution. As is demonstrated by the data in Tables \ref{tablapv1}
and \ref{tablapv2}, the most sensitive observable
appears to be the $Q^2$ projection of $6(-)$, which
  yields a $P$-value of $2.4\times10^{-4}$
for $10^5$ events. Therefore, the
$\mathrm{CP}$-odd differential width $6(-)$ (mainly its $Q^2$
projection) provides a suitable observable for rejecting the SM, since in the
  SM no $\mathrm{CP}$ violation effect is expected
for this decay.  In order to analyze the robustness of the test, we
repeat the procedure with a sample
of $10^6$ events for the scenario in which
  $f_H\eta_P=1.79\,e^{i\pi/4}$. In this case the test seems to loose
its capability of rejection, even for the observable $6(-)$ (see Table
\ref{tablapv3}). The tiny NP contribution in this
  case makes all three $\mathrm{CP}$-odd
observables compatible with zero, at least for $10^6$
  events. This reveals that a larger set of events ($>1\times 10^6$) is needed for
these observables to be useful when the NP contribution is this
small. However, this test can be regarded as an interesting
possibility within the context of the upcoming Super B-factories,
for which a conservative estimate of the expected number of events for
the mode $\tauminusprocess$ gives $\sim 5\times 10^6$, as was already
mentioned in Sec.~\ref{sec5.1}.
\begin{table}[H]
\caption{$P$-values corresponding to the observable $6(-)$ for a NP contribution with $f_H\eta_P=1.79\,e^{i\pi/4}$.}
\label{tablapv3}
\begin{center}
\begin{tabular}{|c|c|c|c|}
\hhline{|====|}\\[-4.5mm]
& \multicolumn{3}{|c|}{$P$-$\mathrm{values}$} \\
\hhline{|====|}
$N_{\mathrm{ev}}/100,000$&
$d\Gamma^{-}_6/dQ^2$&$d\Gamma^{-}_6/ds_1$&$d\Gamma^{-}_6/ds_2$\\[0.5mm]
\hhline{|====|}
$10$ & $0.53$ & $0.93$ & $0.99$ \\[0.5mm]
\hhline{|====|}
\end{tabular}
\end{center}
\end{table}
\subsection{Fitting Procedure}
\label{sec5.3}
\vspace*{-2mm} 
We have performed several fits of the one-dimensional distributions
resulting from the projections of the observables listed in Table
\ref{tabla3} onto $Q^2$, $s_1$ and $s_2$. Only the parameters
appearing linearly in the expressions for the form factors
$F_1$ and $F_2$, namely $A,B,C$,
and $D$, along with the NP parameter, $f_H\eta_P$,
have been taken into account as possible fit parameters, although we
have also tested the possibility of
recovering $N_3$ (which provides information regarding
  the Wess-Zumino contribution) from the fits.\footnote{Although the chosen fitting procedure does
  not take the masses and widths of the resonances as free parameters
  (i.e., these parameters are set to their reference values), we have
  also performed the fits by varying the values for the main
  contributing resonances $K_1(1270)$ and $K_1(1400)$ within the
  uncertainties reported in Ref.~\cite{CLEO}. We have
  observed that these shifts tend to worsen the fits, whereas the uncertainties
  do not change significantly.}
In order to construct the fitting function needed to
apply the least squares method, we write each observable in terms of
the parameters $\underline{\theta}=(A,B,C,D,N_3,f_H\eta^I_P)$\footnote{We note that the fitting procedure introduced in this section could also be applied for the case $\phi_H \neq \pi/2$ by including the parameter $f_H\eta^R_P$ in $\underline{\theta}$.} as
follows, 
\begin{equation}
\label{eq1nueva}
\frac{d\Gamma^{\pm}_i}{dQ^2ds_1ds_2}=\sum_{j}f^{i(\pm)}_j(Q^2,s_1,s_2)\zeta^{i(\pm)}_j(\underline{\theta}),
\end{equation}
where the vectors $\zeta^{i(\pm)}$ depend on the parameters
$\underline{\theta}$ and are listed in Table \ref{tabla1nueva}. By
projecting Eq.~(\ref{eq1nueva}) onto $x\equiv Q^2,s_1,$ or $s_2$, we obtain
the corresponding expected value for the $i$-th projected partial
differential width evaluated for the $k$-th bin of $x$:
\begin{equation}
\label{eq2nueva}
\left( \frac{d\Gamma^{\pm}_i}{dx} \right)_{\mathrm{bin}\,\, k}=\sum_{j}c^{i(\pm)}_{kj}\zeta^{i(\pm)}_j(\underline{\theta}).
\end{equation}
The matrices $c^{i(\pm)}$ in the above
  expression have dimension $N_{\mathrm{bins}}\times
N^{i(\pm)}_{\mathrm{coeff}}$, with $N_{\mathrm{bins}}$
being the number of bins in the $x$ range and
$N^{i(\pm)}_{\mathrm{coeff}}$ being the
number of functions required to express the observable $i(\pm)$ in
terms of the parameters $\underline{\theta}$ appearing in Eq.~(\ref{eq1nueva}).
\begin{table}[H]
\caption{List of the vectors $\zeta^{i(\pm)}$ appearing in Eq.~(\ref{eq2nueva}) expressed in terms of the parameters in $\underline{\theta}$.}
\label{tabla1nueva}
\begin{center}
\begin{tabular}{c|c}
\hline
\hline \\[-0.5cm]
$i(\pm)$ & $\zeta$\\[0.5mm]
\hline
\hline \\[-0.4cm]
$2(+)$ & $(C^2,D^2,CD,A^2,B^2,AB,AC,BC,AD,BD,N^2_3)$ \\[0.7mm]
\hline
$3(+)$ & $(C^2,D^2,CD,A^2,B^2,AB,AC,BC,AD,BD)$\\[0.7mm]
\hline
$4(+)$ &  $(C^2,D^2,CD,A^2,B^2,AB,AC,BC,AD,BD)$\\[0.7mm]
\hline
$5(+)$ &  $(N_3C,N_3D,N_3A,N_3B)$\\[0.7mm]
\hline
$6(+)$ &  $(N_3C,N_3D,N_3A,N_3B)$\\[0.7mm]
\hline
$7(+)$ &  $(CA,CB,DA,DB)$ \\[0.7mm]
\hline
$8(+)$ &  $(N_3C,N_3D,N_3A,N_3B)$\\[0.7mm]
\hline
$9(+)$ &  $(N_3C,N_3D,N_3A,N_3B)$\\[0.7mm]
\hline
$5(-)$ &  $(f_H\eta^I_PC,f_H\eta^I_PD,f_H\eta^I_PA,f_H\eta^I_PB)$\\[0.7mm]
\hline
$6(-)$ &  $(f_H\eta^I_PC,f_H\eta^I_PD,f_H\eta^I_PA,f_H\eta^I_PB)$\\[0.7mm]
\hline
$7(-)$ &  $(f_H\eta^I_PN_3)$\\[0.7mm]
\hline
\hline
\end{tabular}
\end{center}
\end{table}
The different matrices $c^{i(\pm)}$ are obtained by numerical integration of the appropriate function $f^{i(\pm)}_j(Q^2,s_1,s_2)$. With the observables expressed as in Eq.~(\ref{eq2nueva}), we proceed in general to minimize the quantity 
\begin{equation}
\label{eq3nueva}
\chi^2(\underline{\theta}) =\sum_{j=1}^{N_{\mathrm{bins}}}\frac{(y^{\mathrm{sim}}_j-y^{\mathrm{exp}}_j(\underline{\theta}))^2}{\sigma^2_j},
\end{equation}
where the $y^{\mathrm{sim}}_j$ are the values for a given observable extracted from the simulations, the $y^{\mathrm{exp}}_j$ are the corresponding expected values obtained by using the fitting function defined above, and the $\sigma_j$ are the statistical uncertainties associated with the simulation process (see App.~\ref{sec8}). We note that different choices of the parameters in $\underline{\theta}$ with respect to which $\chi^2(\underline{\theta})$ is minimized have been tested. The various resulting fits will be described in the following sections. 
\vspace*{-3mm}
\subsection{Fit Results}
\label{sec5.4}
We present now the results obtained by fitting the $\mathrm{CP}$-odd
as well as the $\mathrm{CP}$-even observables (see Table
\ref{tabla3}). We consider these two sets of
  observables separately.  In the case of the $\mathrm{CP}$-odd
  observables, we regard the NP parameter $f_H\eta_P$ as the unique free parameter and
fix the remaining parameters to their input
values. In the case of the $\mathrm{CP}$-even
  observables we focus on extracting information about the remaining
  parameters, $A,B,C,D$ and $N_3$, from our simulated data.  This
  approach is facilitated by the assumptions mentioned in Sec.\ref{sec3}, namely
that $F_4=f^I_H=0$ and $\phi_H=\pi/2$.
Under these assumptions, the $\mathrm{CP}$-even
observables in Table~\ref{tabla3} do not depend on
the NP contribution, and hence the input value for the parameter
$f_H\eta_P$ is not involved in the analysis of these
observables.\footnote{
Note that $f_H \eta_P$ {\em is} involved in the $\mathrm{CP}$-even
observable ``$1(+)$'' which is not included in Table~\ref{tabla3}. Note
also that, in the more general
  case in which $\phi_H\neq\pi/2$, the $\mathrm{CP}$-even observables
  $5,6$ and $7$ contain NP contributions, but these
are added to the dominant SM
  contribution. By way of contrast, the NP contributions
are dominant for the $\mathrm{CP}$-odd observables in
  the sense that these observables are zero if
  $f_H\eta_P=0$ (since there is no weak phase in $F_4$).}
\vspace*{-3mm}  
\subsubsection{$\mathrm{CP}$-odd observables}
\label{sec5.4.1}
\vspace*{-1mm}
In order to recover the NP parameter $f_H\eta^I_P$ from the
$\mathrm{CP}$-odd observables we perform a least squares fit by fixing
the parameters $A,C$ and $D$ to their input values and setting the
parameter $B$ to zero. The results obtained for two 
data sets (with different numbers of events) for the case
$f_H\eta^I_P=17.9$ are displayed in Tables \ref{tabla5}
and \ref{tabla6}.\footnote{In the
  tables in this and the next sections, the difference between the
  best fit value and the input value for each observable is given in
  units of its respective statistical
  uncertainty, although we use the same symbol $\sigma$
  everywhere.}  The best
 fit value for $f_H\eta^I_P$ is more than $2.5\sigma$ away from zero for all of the $\mathrm{CP}$-odd observables, and is
 more compatible with the input value than with zero. Moreover, this is the case even when
the number of events in the simulation is $5\times 10^ 5$. As was the case for the SM test proposed in the previous section, the observable $6(-)$ appears to be more precise than
the other $\mathrm{CP}$-odd observables (judging by the
smaller statistical uncertainty that it
  yields for the estimated parameter). As can be
seen from the comparison between Tables \ref{tabla5} and \ref{tabla6},
the statistical uncertainties are reduced by
  approximately $50\%$ when the number of events in the simulation
is increased from $5\times 10^5$ to $3\times 10^6$.
\begin{table}[H]
\caption{Best fit values for the parameter $f_H\eta^I_P$ obtained from
  the $\mathrm{CP}$-odd observables with a set of $5\times 10^5$
  events. The input value for the NP
    parameter was set at $f_H\eta^I_P=17.9$. The difference between
  the best fit value and the input value, $|\Delta
  (f_{H}\eta^I_{P})|\equiv
  |\hat{f_{H}\eta^I_{P}}-f_{H}\eta^I_{P}|$, is included.
 }
\label{tabla5}
\begin{center}
\begin{tabular}{|c|c|c||c|c|c||c|c|c|}
\hhline{|=========|}
\multicolumn{9}{|c|}{$N_{\mathrm{ev}}=5\times 10^5$} \\
\hhline{|=========|}\\[-5mm]
%
%
$d\Gamma^{-}_i/dQ^2$& $\hat{f_{H}\eta^{I}_{P}}$ & $|\Delta (f_{H}\eta^I_{P})|$ & $d\Gamma^{-}_i/ds_1$& $\hat{f_{H}\eta^I_{P}}$ & $|\Delta(f_{H}\eta^I_{P})|$ &$d\Gamma^{-}_i/ds_2$& $\hat{f_{H}\eta^I_{P}}$ & $|\Delta(f_{H}\eta^I_{P})|$\\[0.5mm]
\hhline{|=========|} \\[-0.5cm]
$5$ & $28 \pm 11$ & $0.9\sigma$ & $5$ & $21 \pm 8$ & $0.4\sigma$ & $5$ & $19 \pm 5$ & $0.2\sigma$ \\[0.6mm]
\hline
$6$ & $17 \pm 1$ & $0.9\sigma$ & $6$ & $18 \pm 1$ & $0.1\sigma$ & $6$ & $17 \pm 1$ & $0.9\sigma$ \\[0.6mm]
\hline
$7$ & $20 \pm 4$ & $0.5\sigma$ & $7$ & $17 \pm 4$ & $0.2\sigma$ & $7$ & $19 \pm 4$ & $0.3\sigma$\\[0.6mm]
\hhline{|=========|}
\end{tabular}
\end{center}
\end{table}
\begin{table}[H]
\caption{Best fit values for the parameter $f_H\eta^I_P$ obtained from the $\mathrm{CP}$-odd observables with a set of $3\times 10^6$ simulated events . The difference $|\Delta (f_{H}\eta^I_{P})|\equiv |\hat{f_{H}\eta^I_{P}}-f_{H}\eta^I_{P}|$ is included.}
\label{tabla6}
\begin{center}
\begin{tabular}{|c|c|c||c|c|c||c|c|c|}
\hhline{|=========|}
\multicolumn{9}{|c|}{$N_{\mathrm{ev}}=3\times 10^6$} \\
\hhline{|=========|}\\[-4.9mm]
$d\Gamma^{-}_i/dQ^2$& $\hat{f_{H}\eta^I_{P}}$ & $|\Delta (f_{H}\eta^I_{P})|$ & $d\Gamma^{-}_i/ds_1$& $\hat{f_{H}\eta^I_{P}}$ & $|\Delta (f_{H}\eta^I_{P})|$ & $d\Gamma^{-}_i/ds_2$& $\hat{f_{H}\eta^I_{P}}$ & $|\Delta (f_{H}\eta^I_{P})|$\\[0.5mm]
\hhline{|=========|} \\[-0.5cm]
$5$ & $18 \pm 5$ & $0.02\sigma$ & $5$ & $22 \pm 3$ & $1.4\sigma$ & $5$ & $18 \pm 2$ & $0.05\sigma$ \\[0.6mm]
\hline
$6$ & $17.6 \pm 0.4$ & $0.8\sigma$ & $6$ & $18.0 \pm 0.5$ & $0.2\sigma$ & $6$ & $17.4 \pm 0.5$ & $1.0\sigma$ \\[0.6mm]
\hline
$7$ & $17 \pm 2$ & $0.5\sigma$ & $7$ & $15 \pm 2$ & $1.5\sigma$ & $7$ & $17 \pm 2$ & $0.5\sigma$\\[0.6mm]
\hhline{|=========|}
\end{tabular}
\end{center}
\end{table}
We have also performed a least squares fit using the set of
$10^6$ events with $f_H\eta_P=1.79\,e^{i\pi/4}$. In
this case the best values obtained from the fit to the observables
$5(-)$ and $7(-)$ become compatible with zero and have large
statistical uncertainties, whereas the observable $6(-)$ is still the
most precise one, giving best fit values
that are more than $2\sigma$ away from zero and that recover the
input value $f_H\eta^I_P=1.79\,\sin(\pi/4)\simeq
  1.27$ even though the uncertainties
are larger than those we obtain with $f_H\eta_P=17.9\,e^{i\pi/2}$ using a set of $10^6$ events.\footnote{For the case with $f_H\eta_P=17.9\,e^{i\pi/2}$ we only display results obtained using $5\times 10^5$ and $3\times 10^6$ events, although we have also  performed similar fits using sets of events of different sizes.}
The results for the three projections of
the observable $6(-)$ are shown in Table \ref{tabla7}.
\begin{table}[H]
\vspace*{-2mm}
\caption{Best fit values for the parameter $f_H\eta^I_P$ obtained from the observable $6(-)$ by using a set of $10^6$ simulated events with an input value $f_H\eta_P=1.79\,e^{i\pi/4}$ (so that
$f_H\eta_P^I\simeq 1.27$).}
\label{tabla7}
\begin{center}
\begin{tabular}{|c|c|c||c|c|c||c|c|c|}
\hhline{|=========|}
\multicolumn{9}{|c|}{$N_{\mathrm{ev}}=1\times 10^6$} \\
\hhline{|=========|}\\[-4.9mm]
$d\Gamma^{-}_i/dQ^2$& $\hat{f_{H}\eta^I_{P}}$ & $|\Delta(f_{H}\eta^I_{P})|$ & $d\Gamma^{-}_i/ds_1$& $\hat{f_{H}\eta^I_{P}}$ & $|\Delta(f_{H}\eta^I_{P})|$ & $d\Gamma^{-}_i/ds_2$& $\hat{f_{H}\eta^I_{P}}$ & $|\Delta(f_{H}\eta^I_{P})|$\\[0.5mm]
\hhline{|=========|} \\[-0.5cm]
$6$ & $1.9 \pm 0.6$ & $1.1\sigma$ & $6$ & $1.8 \pm 0.8$ & $0.7\sigma$ & $6$ & $1.7 \pm 0.8$ & $0.5\sigma$\\[0.6mm]
\hhline{|=========|}
\end{tabular}
\end{center}
\end{table}

Both the results obtained from the least squares fit and the SM test
indicate the utility of using the observable $6(-)$ as
  a tool for investigating $\mathrm{CP}$-odd NP effects. On the one hand, the SM test shows
this observable's power to reject the SM
hypothesis if there is actually a $\mathrm{CP}$-violating
  contribution; on the other hand, the least squares fit demonstrates
 how this observable can be used to recover the input value of
  the NP parameter.  It is interesting to consider why the
  $6(-)$ observable is so much more sensitive to $\mathrm{CP}$
  violation than are the other two $\mathrm{CP}$-odd observables that
  we have considered.  This sensitivity
arises from the dependence of the $\mathrm{CP}$-odd
observables on the quantities $B_i$. As is evident
in Table \ref{tabla3}, the $7(-)$ observable is
doubly suppressed due to the smallness of the W-Z and the NP
contributions. Similarly, comparison of the $5(-)$ and $6(-)$ observables
  indicates that the latter exhibits a larger magnitude (and hence greater
  sensitivity to NP) because it depends on the quantity $B_1$,
whereas the former depends on $B_2$; numerical study
  has shown that the magnitude of $B_1$ tends to be larger than that
  of $B_2$ within the allowed ranges of $Q^2,s_1$ and $s_2$.\par

\begin{figure}[H]
\centering
\subfloat[][]{\includegraphics[width=0.45\textwidth,height=0.33\textwidth]
{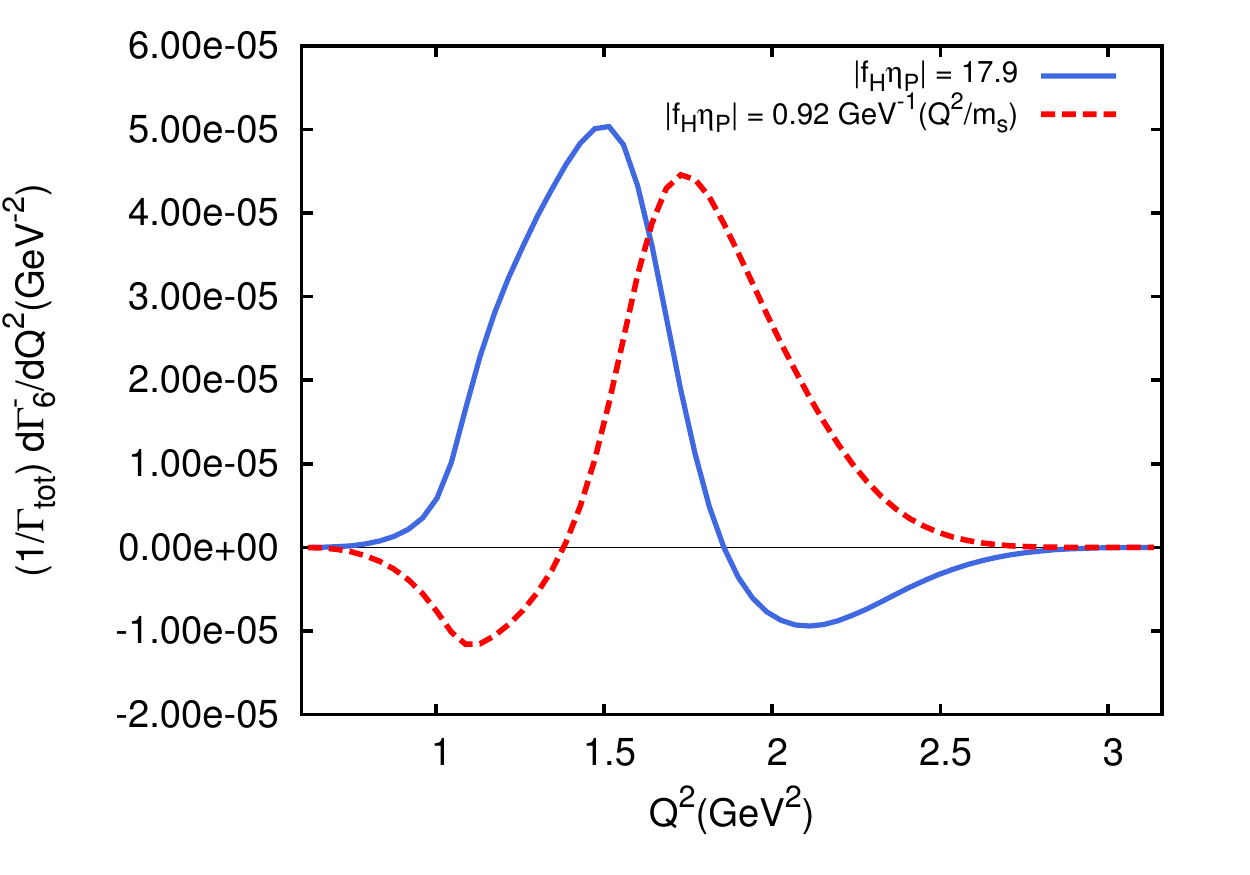} 
\label{fig6odda}}
\subfloat[][]{\includegraphics[width=0.45\textwidth,height=0.33\textwidth]
{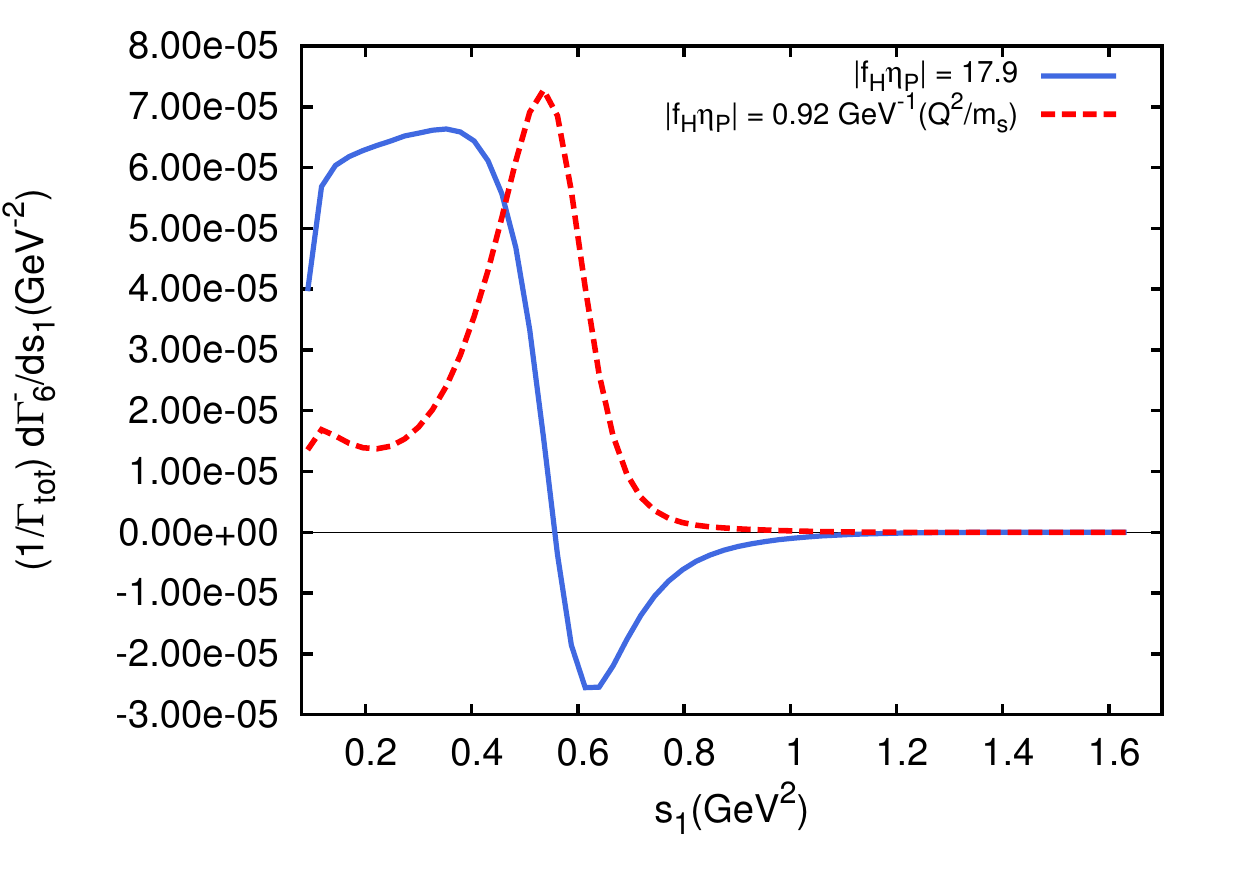} 
\label{fig6oddb}}
\quad
\subfloat[][]{\centering \includegraphics[width=0.45\textwidth,height=0.33\textwidth]
{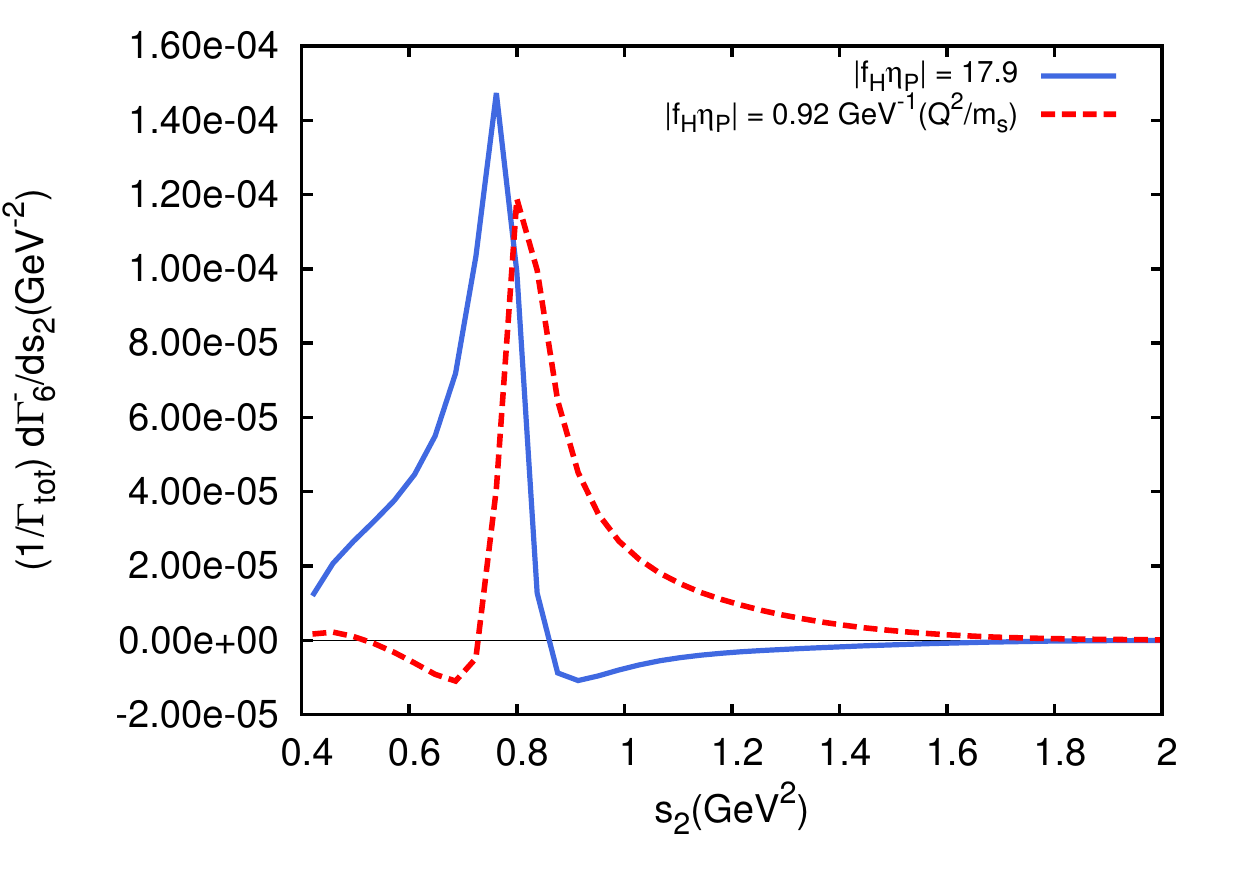} 
\label{fig6oddc}}
\caption{Plots of the distributions obtained from the observable
  $6(-)$ for $|f_H\eta_P|=17.9$ (blue solid line) and $|f_H\eta_P|=0.92\,\mathrm{GeV}^{-1}(Q^2/m_s)$ (red dashed line). In the panels (a),(b) and (c) the projections onto $Q^2,s_1$
  and $s_2$ are displayed, respectively. The distributions are normalized to the total width of the $\tau$.}
\label{fig6odd}
\end{figure}

The above results are based on the assumption that
  $f_H$ has no $Q^2,s_1$ or $s_2$ dependence. It is important to note,
  however, that a non-trivial dependence on the kinematical variables
  could appear due to the presence of final state interactions.  The
  functional form of $f_H$ is unknown at present.  Having said this,
  it is instructive to adopt a simple functional form for $f_H$ in
  order to test how the $6(-)$ distributions are modified.  For the
  purpose of illustration, let us reconsider the expression for $f_H$
  derived from the quark equations of motion, $f_H\sim(Q^2/m_s)F_4$,
  where $F_4$ is assumed to be a constant. In order to set a reference
  value for $|F_4|$ the expression derived in Ref.~\cite{Decker}
  within the context of Chiral Perturbation Theory has been used. A
  numerical analysis similar to that discussed in Sec.~\ref{sec3}
  gives $\langle |F_4|\rangle\sim 0.54\,\mathrm{GeV}^{-1}$ and
  $\mathcal{O}(\langle \mathrm{Im}(F_4)\rangle) <\mathcal{O}( \langle
  \mathrm{Re}(F_4)\rangle)$.  We set
  $F_4=0.54\,\mathrm{GeV}^{-1}$ and add a normalization
  factor in the expression for $f_H$, $\mathcal{N}$, so that the
  experimental uncertainty of the branching ratio is again saturated
  by the NP contribution. By taking $\phi_H=\pi/2$ we find the value
  $\mathcal{N}|\eta_P|=1.71$. Hence, $|f_H
  \eta_P|=\mathcal{N}(Q^2/m_s)F_4|\eta_P|=1.71\times
  0.54\,\mathrm{GeV}^{-1}
  (Q^2/m_s)=0.92\,\mathrm{GeV}^{-1}(Q^2/m_s)$. Figure~\ref{fig6odd} shows plots of the 
  $6(-)$ distributions for the case $f_H\eta^I_P=17.9$ (blue solid line) along with the specific case
  presented above in which $f_H$ depends linearly on $Q^2$ (red
  dashed line).
  \vspace*{-4mm}
  \noindent These distributions have been obtained
  numerically and normalized to the total width of the
  $\tau$ ($\Gamma_{\mathrm{tot}}$). As can be seen from the plots, the
  $6(-)$ distributions arising from the two approaches are
  comparable. On the one hand, the order of magnitude of each
  projection remains the same in both cases. On the other hand, the
  maxima of the distributions do not change significantly from one
  approach to the other. Based on these facts, it would be reasonable
  to expect that the number of events needed for recovering the NP
  parameter from the $6(-)$ distributions in the case
  $|f_H\eta_P|=17.9$ would also be enough for the case
  $|f_H\eta_P|\propto Q^2$. In this sense, the presence of a linear
  $Q^2$ dependence in $f_H$ should not spoil the sensitivity of the
  $6(-)$ distributions to the NP contribution with respect to the case
  in which $|f_H|$ is assumed to be a flat function. Hence, this
  specific case shows that the proposed observables
  could be useful even when there is a non-trivial dependence of
  $|f_H|$ on $Q^2, s_1$ and $s_2$.
\subsubsection{$\mathrm{CP}$-even observables}
\label{sec5.4.2}
In this section we focus on $\mathrm{CP}$-even
observables. We will discuss the results arising from the observables
$2(+)-9(+)$ and then, separately, those arising from the $1(+)$ distribution,
 due to its preferential treatment in previous analyses
\cite{CLEO,Belle}.\par 
In order to test the power of the method, we first performed a fit
with the parameters $A,C,D$ and $N_3$ unconstrained and $B$ set to
zero. In this case, we observe that the correlation between the
parameters, as well as the standard deviations, are very large and the
outputs of the fit for the different parameters are far away from the
input values. To address these issues, we have adopted a modified fit procedure,
in which the parameters $C$ and $D$ are constrained by the branching fractions into
the $K^*\pi$ final state from the $K_1(1270)$ and $K_1(1400)$,
respectively (see Ref.~\cite{CLEO}). In addition, we keep the
parameters $B$ and $N_3$ fixed to their input values, $B=0$ and
$N_3=1.4696$, respectively. Accordingly, we have minimized the
distributions only with respect to the parameter $A$. The results of
the fit for $3\times 10^ 6$ events are tabulated in Table
\ref{tabla8}.
\begin{table}[H]
\caption{Fit results for the parameter $A$ obtained
  from the $\mathrm{CP}$-even observables $2(+)-7(+)$
  using a sample of $3\times 10^ 6$ simulated events.
  The input value for the simulation was taken to be
   $A=0.944$. The difference $|\Delta
  A|\equiv |\hat{A}-A|$ is also displayed.}
\label{tabla8}
\begin{center}
\begin{tabular}{|c|c|c||c|c|c||c|c|c|}
\hhline{|=========|}
\multicolumn{9}{|c|}{$N_{\mathrm{ev}}=3\times 10^6$} \\
\hhline{|=========|}\\[-4.9mm]
$d\Gamma^{+}_i/dQ^2$& $\hat{A}$ & $|\Delta A|$ & $d\Gamma^{+}_i/ds_1$& $\hat{A}$ & $|\Delta A|$ & $d\Gamma^{+}_i/ds_2$& $\hat{A}$ & $|\Delta A|$  \\[0.5mm]
\hhline{|=========|} \\[-0.5cm]
$2$ & $0.95 \pm 0.01$ & $0.6\sigma$ & $2$ & $0.94 \pm 0.01$ & $0.4\sigma$ & $2$ & $0.94 \pm 0.02$ & $0.2\sigma$ \\[0.6mm]
\hline
$3$ & $0.92 \pm 0.01$ & $2.4\sigma$ & $3$ & $0.91 \pm 0.01$ & $3.4\sigma$ & $3$ & $0.91 \pm 0.02$ & $1.7\sigma$ \\[0.6mm]
\hline
$4$ & $0.93 \pm 0.02$ & $0.7\sigma$ & $4$ & $0.94 \pm 0.01$ & $0.4\sigma$ & $4$ & $0.94 \pm 0.02$ & $0.2\sigma$  \\[0.6mm]
\hline
$5$ & $0.949 \pm 0.008$ & $0.6\sigma$ & $5$ & $0.947 \pm 0.006$ & $0.6\sigma$ & $5$ & $0.942 \pm 0.006$ & $0.3\sigma$  \\[0.6mm]
\hline
$6$ & $0.91 \pm 0.03$ & $1.1\sigma$ & $6$ & $0.92 \pm 0.02$ & $1.2\sigma$ & $6$ & $0.92 \pm 0.02$ & $1.2\sigma$ \\[0.6mm]
\hline
$7$ & $0.94 \pm 0.01$ & $0.4\sigma$ & $7$ & $0.942 \pm 0.008$ & $0.3\sigma$ & $7$ & $0.948 \pm 0.005$ & $0.8\sigma$\\[0.6mm]
\hhline{|=========|}
\end{tabular}
\end{center} 
\end{table}
Before we discuss the results
  in Table~\ref{tabla8}, we note that the $8(+)$ and $9(+)$ distributions extracted from the set of $3\times 10^6$ simulated events are consistent with zero to within their statistical uncertainties (which are determined using Eq.~(\ref{Apeq2})). As a result, no conclusive information can be obtained from these observables with this number of events. For this reason we do not include results from these observables in the table. Turning now to the observables $2(+)-7(+)$, we notice that for these
observables the input value is recovered in all cases with uncertainties smaller than $3\%$;
furthermore, the three projections of $5(+)$ and the $s_{1,2}$ projections of $7(+)$ are the
most precise, with uncertainties smaller than $1\%$.\par
We turn now to a consideration of the observable $1(+)$.
All of the projections of this observable are
positive distributions that are more than two orders
of magnitude larger than those arising
from the other $\mathrm{CP}$-even observables.
Since the absolute statistical uncertainties are
  similar for all of the $\mathrm{CP}$-even distributions, the $1(+)$
  distributions end up having considerably reduced relative
  statistical uncertainties compared to those for the other
  $\mathrm{CP}$-even distributions. Therefore, we have analyzed
this distribution in a different manner, allowing $A, N_3$ and $f_H\eta_P$ 
to float as free parameters.  Although the best fit point obtained
from the fit to the $1(+)$ distribution is in good agreement with the
corresponding input values, and the
standard deviations are smaller than those associated with the other
observables, there are certain disadvantages in the use of this
distribution for extracting the value of $f_H\eta_P$. First of all, it
is important to note that the fact that the distribution appears to be
sensitive to the NP contribution arises exclusively from the
input value that we have used for the NP parameter. More precisely, as
outlined above, the NP parameter has been set to a
  value such that it saturates the experimental uncertainty, which
includes both statistical and systematic sources.
This experimental uncertainty is
higher than the uncertainty associated with extracting the distributions from the simulations,  which is purely statistical. Moreover, the statistical uncertainty that we have used in our analysis is smaller than the statistical uncertainties in the experiments since we are using a larger number of events for our simulation. Therefore, in our analysis, the NP contribution exceeds the statistical uncertainties of the simulated $1(+)$ distribution,
leading to a best fit value for $f_H\eta_P$ essentially incompatible
with zero. This observation is supported
by the fact that when we carry out the same fit using the
set of events simulated with $f_H\eta_P=1.79\,e^{i\pi/4}$, we obtain a
best fit value in agreement with zero. Moreover,
the computation of the correlation matrix for both sets of events
shows that there are significant correlations between the fit
parameters. Furthermore, the
least squares function that we minimize exhibits
several local minima that are not far enough from the global minimum
to distinguish them if the precise input values are not known
beforehand. It is worth
noting that this sort of problem is absent when we fit the $\mathrm{CP}$-odd observables in
order to obtain the single NP parameter.\footnote{This could arise from the fact that, for the observable $1(+)$,  the $\chi^2$ is a quartic function of the input parameters, whereas for the $\mathrm{CP}$-odd observables it is a quadratic function of the NP parameter.} Lastly, note that under the assumptions used in
this work, one would not be able to extract any information about the
NP weak phase from the analysis of the $1(+)$ distribution because its
dependence on the NP parameter enters as
the squared modulus of $B_4$ and $\overline{B}_4$, which are
proportional to $|\eta_P|$ under our assumption that $F_4=0$ (see
Eqs.~(\ref{eq11}), (\ref{eq12}) and (\ref{eq15})). Even if
$F_4\neq 0$, the dependence on the NP parameter would be mixed in a
complicated way with the dependence on the SM scalar form factor
$F_4$, preventing their disentanglement. We remark that the inability to distinguish the NP
contribution from the SM contribution is common to all the
$\mathrm{CP}$-even observables, while it is absent in the case of the
$\mathrm{CP}$-odd observables.
\hfill\break
\par
Several of the $\mathrm{CP}$-even observables are in principle
sensitive to the parameter $N_3$ (which fixes the contribution of the
anomalous Wess-Zumino term). However, as was noted above, the 
$8(+)$ and $9(+)$ distributions are consistent with zero, even with the 
maximum number of events that we have simulated. This spoils the sensitivity of these 
observables to the parameter $N_3$. An alternative is to use the observables
$5(+)$ and/or $6(+)$ with the parameters $A,C$ and $D$ fixed to their
input values.  With these parameters fixed in this way, the $5(+)$ and $6(+)$ 
distributions depend only on
$N_3$. Of course, when experimental data is used instead of simulated
events, the input values will be unknown.  In this case, one could use the other
observables to estimate the parameter $A$ first; then 
$C$ and $D$ could be obtained by applying
constraints arising from the tabulated branching fractions of the $K_1$ resonances (see Eqs. (8)-(10) in Ref.~\cite{CLEO}). The results for $N_3$ obtained from the $5(+)$ and
$6(+)$ distributions are shown in Table \ref{tabla9} for a simulation using $3\times 10^ 6$ events.  Both observables allow one to recover the parameter $N_3$.  The observable $5(+)$, however, is the more precise of
the two; its
uncertainties are smaller than $4\%$, while those associated with 
the $6(+)$ distribution are of order $15\%$. Hence, the observable $5(+)$ appears to be the most
appropriate observable for implementing the proposed strategy to extract information about the anomalous Wess-Zumino contribution.
\begin{table}[H]
\caption{Results for $N_3$ from fits to the $5(+)$ and $6(+)$ distributions with a set of $3\times 10^6$ simulated events. The fit has been performed by fixing the parameters $A,C$ and $D$ to their input values.
The input value for $N_3$ was $1.4696$.}
\label{tabla9}
\begin{center}
\begin{tabular}{|c|c|c||c|c|c||c|c|c|}
\hhline{|=========|}
\multicolumn{9}{|c|}{$N_{\mathrm{ev}}=3\times 10^6$} \\
\hhline{|=========|}\\[-4.8mm]
$d\Gamma^{+}_i/dQ^2$& $\hat{N_3}$ & $|\Delta N_3|$ & $d\Gamma^{+}_i/ds_1$& $\hat{N_3}$ & $|\Delta N_3|$ & $d\Gamma^{+}_i/ds_2$& $\hat{N_3}$ & $|\Delta N_3|$\\[0.5mm]
\hhline{|=========|} \\[-0.5cm]
$5$ & $1.45 \pm 0.05$ & $0.4\sigma$ & $5$ & $1.45 \pm 0.05$ & $0.4\sigma$ & $5$ & $1.49 \pm 0.05$ & $0.4\sigma$ \\[0.6mm]
\hline
$6$ & $1.3 \pm 0.2$ & $0.9\sigma$ & $6$ & $1.5 \pm 0.2$ & $0.2\sigma$ & $6$ & $1.5 \pm 0.2$ & $0.2\sigma$\\[0.6mm]
\hhline{|=========|}
\end{tabular}
\end{center} 
\end{table}

We conclude this section by summarizing, in Table
  \ref{tabla10}, the main results obtained for the $6(-)$ observable.
  Of the various observables proposed in this work, the $6(-)$
  distribution shows the most promise for detecting $\mathrm{CP}$-odd
  NP effects in $\tauprocess$.

\renewcommand{\arraystretch}{1.5}
\begin{table}[H]
\caption{Main results for the $6(-)$ observable obtained in Sec.~\ref{sec5} by using various sets of simulated events with $|f_H\eta_P|=17.9$.}
\label{tabla10}
\begin{center}
\begin{tabular}{|c||c|c||c|c|}
\hhline{|=====|}
\multirow{2}{*}{ Distribution } & \multicolumn{2}{ c|| }{ ~\bf{SM hypothesis test}~ } & \multicolumn{2}{ c| }{ ~\bf{Least Squares fit}~ }\\ 
\cline{2-5}
& $ \,\,N_{\mathrm{ev}}\,\, $ & $ P$-$\mathrm{value} $ & $\,\, N_{\mathrm{ev}}\,\, $ & ~Fit value for $f_H\eta^I_P$~ \\
\hhline{|=====|}
$d\Gamma^{-}_6/dQ^2$ & ~~$ 10^5 $~~ & $ 0.000024 $ & ~~$ 3\times 10^6 $~~ & ~~$ 17.6 \pm 0.4 $~~ \\[0.2mm]
\hline 
$d\Gamma^{-}_6/ds_1$ & ~~$ 10^5 $~~ & $ 0.0076 $ & ~~$ 3\times 10^6 $~~ & ~~$ 18.0 \pm 0.5 $~~ \\[0.2mm] 
\hline
$d\Gamma^{-}_6/ds_2$ & ~~$ 10^5 $~~ & $ 0.013 $ & ~~$ 3\times 10^6 $~~ & ~~$ 17.4 \pm 0.5 $~~ \\[0.2mm] 
\hhline{|=====|}
\end{tabular}
\end{center}
\end{table}
\renewcommand{\arraystretch}{1.0}
\section{{\large $\tau\rightarrow K\pi\pi\nu_{\tau}$} within the aligned 2HDM}
\label{sec6}
So far we have analyzed the decay $\tauprocess$ in a model-independent
framework, in which the NP effects are incorporated by adding the
contribution of a charged scalar boson that couples to fermions in a ``non-standard'' manner
(i.e., the couplings are not suppressed by
the masses of the light quarks \cite{KA}). In this
  section we consider the proposed analysis in the context of a
  particular model of NP. Many NP models extend the
SM scalar sector by adding a second scalar doublet so that the scalar spectrum contains a charged boson. A particular example of such a model is the so-called
aligned two-Higgs-doublet model (A2HDM) \cite{Pich1}. In the A2HDM, 
an alignment between Yukawa coupling matrices leads
to the elimination of the non-diagonal neutral couplings
  that would lead to tree-level flavour-changing neutral currents.
 \par The Yukawa Lagrangian
corresponding to the charged Higgs boson in the A2HDM
can be written in terms of the fermion mass eigenstates as
\cite{Pich1,Pich2}
\begin{equation}
\label{eq30}
\mathcal{L}^{H^{\pm}}_Y= -\frac{\sqrt{2}}{v}H^+\{\overline{u}[\varsigma_d V M_d \mathcal{P}_R-\varsigma_u M_u V \mathcal{P}_L]d + \varsigma_l\overline{\nu} M_l \mathcal{P}_R l\} + \mathrm{h.c.},
\end{equation}
where $M_{u,d}$ are the diagonal mass matrices, $V$ is the Cabibbo-Kobayashi-Maskawa (CKM) matrix,
$v$ is the Higgs vacuum expectation value and
$\mathcal{P}_{R,L}\equiv \frac{1\pm \gamma_5}{2}$ are the chirality
projection operators.  The proportionality parameters $\varsigma_f~
(f=u,d,l)$ are arbitrary complex numbers and give rise to new sources
of $\mathrm{CP}$ violation.

From Eq.~(\ref{eq30}) we see that within the A2HDM the effective
couplings $g^{q_uq_dl}_L$ and $g^{q_uq_dl}_R$ appearing in the
corresponding effective Hamiltonian are given by \cite{Pich1}
\begin{equation}
\label{eq31}
g^{q_uq_dl}_L=\varsigma_u\varsigma^*_l\frac{m_{q_u}m_l}{M^2_{H^{\pm}}},\qquad\quad g^{q_uq_dl}_R=-\varsigma_d\varsigma^*_l\frac{m_{q_d}m_l}{M^2_{H^{\pm}}}.
\end{equation}
Moreover, given the three-family universality of the proportionality parameters $\varsigma_f$, the following relations are satisfied,
\begin{equation}
\label{eq32}
\frac{g^{q_u q_d l}_L}{g^{q'_u q'_d l'}_L}=\frac{m_{q_u}m_l}{m_{q'_u}m_{l'}},\qquad\quad 
\frac{g^{q_u q_d l}_R}{g^{q'_u q'_d l'}_R}=\frac{m_{q_d}m_l}{m_{q'_d}m_{l'}}.
\end{equation}
In our case, the relations between the couplings $\eta_{P,S}$ defined
in Eq.~(\ref{eq2}) and those introduced in Eq.~(\ref{eq31}) are given
by
\begin{equation}
\label{eq33}
\frac{\eta^*_S+\eta^*_P}{2}\,=\,g^{us\tau}_L\,\dot{=}\,\varsigma_u\varsigma^*_l\frac{m_{u}m_{\tau}}{M^2_{H^{\pm}}}\,\quad \mbox{and}\,\quad \frac{\eta^*_S-\eta^*_P}{2}\,=\,g^{us\tau}_R\,\dot{=}-\varsigma_d\varsigma^*_l\frac{m_{s}m_{\tau}}{M^2_{H^{\pm}}},
\end{equation} 
where the last equalities hold only within the A2HDM. Owing to the $m_u$
suppression, $g^{us\tau}_L$ can be neglected and the relations in Eq.~(\ref{eq33}) reduce to
\begin{equation}
\label{eq34}
\eta_P\,=\,-g^{us\tau*}_R\,\dot{=}\,\varsigma^*_d\varsigma_l\frac{m_{s}m_{\tau}}{M^2_{H^{\pm}}}.
\end{equation}
The above expression, along with the second relation in Eq.~(\ref{eq32}), imply that observables from other systems involving the couplings $g^{q_u q_d l}_R$ will provide constraints for the pseudoscalar coupling $\eta_P$, which can be used
in turn to obtain predictions for the observables proposed in
Sec.~\ref{sec3}. In this case, the observables we
  have proposed could be useful for testing the
A2HDM.\par
\begin{figure}[H]
\centering
\includegraphics[width=0.57\textwidth,height=0.43\textwidth]{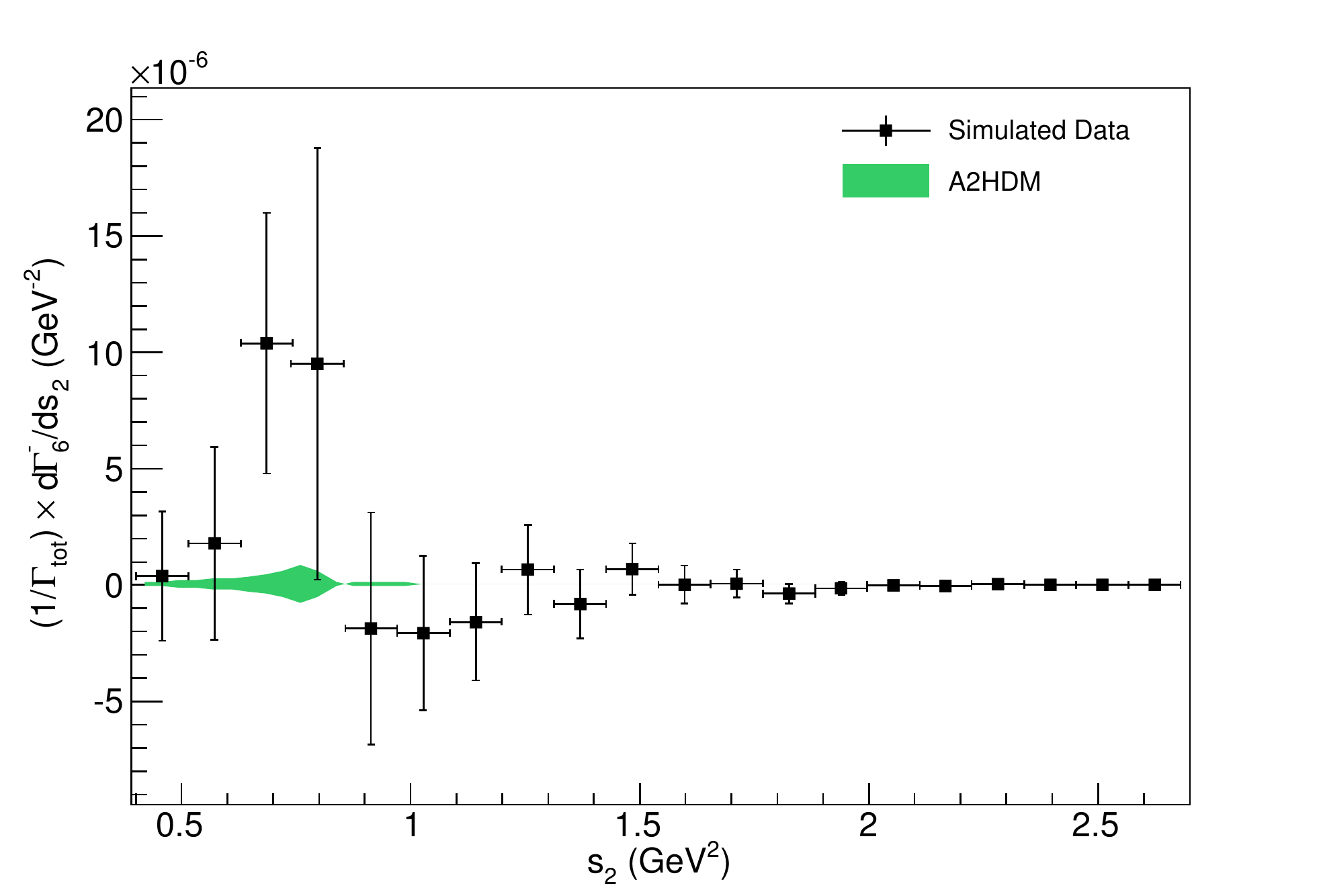}
\caption{Projection onto $s_2$ of the observable $6(-)$ extracted from a set of $10^6$ events along with the corresponding allowed region within
  the A2HDM. The data in the simulation corresponds to the NP parameter choice
$f_H\eta_P=1.79\,e^{i\pi/4}$.
Note that the plot of the allowed region assumes that the parameters associated with the form factors ($A$, $B$, etc.) have zero uncertainty.}
\label{fig2}
\end{figure}

 Let us now consider an example that
  will illustrate how outside constraints can be used to make testable
  predictions in $\tauprocess$. In this example we will focus on the observable $6(-)$, which happens to be much more sensitive to $\mathrm{CP}$ violation than the other proposed observables, as was discussed in Sec.~\ref{sec5.4.1}. The phenomenology derived from the A2HDM has been studied
extensively (see for example Refs.~\cite{Pich1,Branco}). In particular, the
constraints obtained by combining the information from various
semileptonic and leptonic decays have been discussed in
Refs.~\cite{Pich1,Pich2}. Hence, guided by Ref.~\cite{Pich2}, and
assuming that $1<|f_H|<10$ and that $f^I_H=0$, we
derive the (model-dependent) constraints
$-0.01<f_H\eta^{R,I}_P<0.01$. It should be
noted that in this case we are considering an arbitrary weak phase
$\phi_H$, in contrast with our analysis in Sec.~\ref{sec5}, in which the analysis was
restricted to $\phi_H=\pi/2, \pi/4$. In order to test the A2HDM, the $6(-)$ distributions extracted from the data can be compared to the corresponding allowed region arising from the very restrictive bound mentioned above. Since we are using simulated events instead of experimental data, we will make use of the $6(-)$ distributions extracted from our simulations.  In particular, we will use
the distributions associated with the NP parameter choice 
$f_H\eta_P=1.79\,e^{i\pi/4}$, instead of those associated
with $f_H\eta_P=17.9\,e^{i\pi/2}$, since the former parameter choice
is closer to the range obtained from the A2HDM.
In addition, we note that this parameter choice is compatible with the constraints derived in a model-independent manner from the decay $\tau\rightarrow K\nu_{\tau}$ (assuming that $f^I_H=0$ and that $1<|f_H|<10$), regardless of whether one uses the quark or meson mass to determine the bound. The projection onto $s_2$ of the observable $6(-)$ is displayed in Fig.~\ref{fig2} along with the prediction derived from the A2HDM. We consider only the $s_2$ projection because it tends to have the largest magnitude
for this observable.
Inspection of Fig.~\ref{fig2} reveals that the distribution lies
outside the A2HDM prediction only in the 3th and 4th bins, with the
deviations being smaller than $2\sigma$ and almost $1\sigma$,
respectively. However, as was already shown in Sec.\ref{sec5.4.1},
when we perform a least squares fit to this distribution with
$f_H\eta^I_P$ as the unique free parameter, we obtain the value $1.7
\pm 0.8$ (see Table \ref{tabla7}), which is more than $2\sigma$ away
from the range allowed for this parameter within the A2HDM
($|f_H\eta^I_P|<0.01$). Although such a deviation would cast doubt on
the A2HDM in an experimental setting, it would not be enough to completely reject the
model. Thus, for a NP parameter $f_H\eta^I_P$ two orders of magnitude
above the range predicted by the A2HDM, more than $10^6$
events would be needed for the observable $6(-)$ to be useful in
probing this model.  A similar observation holds for the case of the
SM, since in that case the $6(-)$ distribution is simply
zero and is thus contained within the
range allowed for the A2HDM.  In fact, the situation here is similar
to the situation that was considered in Secs.~\ref{sec5.2} and
\ref{sec5.4.1}, where it was noted that more than $10^6$ events were
required to use the $6(-)$ distribution as a tool for distinguishing
between the SM and a NP scenario with $|f_H\eta_P|=1.79$.

Finally, we emphasize that the allowed region indicated in Fig.~\ref{fig2} 
assumes that the pseudoscalar form factor $f_H$ is a constant function
of the phase space variables and that its imaginary part is zero.
In order to perform a more realistic study of
the A2HDM within the context of the observables discussed in this work, these assumptions
would need to be tested carefully. In Sec.~\ref{sec7} we comment on some
possibilities for testing these assumptions.  

\section{Test of assumptions}
\label{sec7}
As has been mentioned in previous
sections, various assumptions have
been made while performing the analysis in this work. Some of
these assumptions could in
principle be tested by using the proposed
observables. In this section we describe how one could test two assumptions that have been made
regarding the pseudoscalar form factor $f_H$; namely, that it is a flat function of $Q^2,s_1$ and $s_2$, 
and that it does not contain strong phases (i.e., that $f^I_H$ is zero).\par From the observables
$5(-)$ and $6(-)$ in Table \ref{tabla3} we have the following
relations
\begin{equation}
\label{eq35}
\frac{d\Gamma^-_5 }{dQ^2ds_1ds_2}=\left(\frac{2}{3}A(Q^2)\langle\overline{K}_2\rangle\frac{\sqrt{Q^2}}{m_{\tau}}B^R_2\right)\!f^I_H\eta^I_P-\left(\frac{2}{3}A(Q^2)\langle\overline{K}_2\rangle\frac{\sqrt{Q^2}}{m_{\tau}}B^I_2\right)\!f^R_H\eta^I_P\quad\,\,\hspace*{0.96mm}
\end{equation} 
\begin{equation}
\label{eq36}
\frac{d\Gamma^-_6 }{dQ^2ds_1ds_2}=-\left(\frac{2}{3}A(Q^2)\langle\overline{K}_2\rangle\frac{\sqrt{Q^2}}{m_{\tau}}B^R_1\right)\!f^I_H\eta^I_P+\left(\frac{2}{3}A(Q^2)\langle\overline{K}_2\rangle\frac{\sqrt{Q^2}}{m_{\tau}}B^I_1\right)\!f^R_H\eta^I_P\,,
\vspace*{2mm}
\end{equation} 
where we recall that the quantities $B_1,B_2$ and $\langle\overline{K}_2\rangle$ depend on the kinematical variables $Q^2,s_1$ and $s_2$. By projecting Eqs.~(\ref{eq35}) and (\ref{eq36}) onto $x\equiv Q^2,s_1,s_2$ we can form a $2\times2$ matrix equation 
\begin{equation}
\label{eq37}
\left(\begin{array}{c}
d\Gamma^-_5/dx \\[1mm]
d\Gamma^-_6/dx 
\end{array}\right)=\left(\begin{array}{rr}
a_1 & -b_1 \\[1mm]
-a_2 & b_2
\end{array}\right)\left(\begin{array}{c}
f^I_H \eta^I_P \\[1mm]
f^R_H \eta^I_P
\end{array}\right),
\end{equation}
where the quantities $a_1$ and $b_1$ are the projections onto $x$ of the two
functions appearing inside the parentheses in Eq.~(\ref{eq35}), while
$a_2$ and $b_2$ arise from the two functions in Eq.~(\ref{eq36}). Of course,
these quantities are functions of $x$. Also, we note that we need to
assume that $f_H$ has no dependence on the kinematical variables other than $x$ in
order to derive Eq.~(\ref{eq37}). By inverting
Eq.~(\ref{eq37}) we obtain the relations
\begin{equation}
\label{eq38}
f^I_H\eta^I_P=\frac{1}{a_1b_2-a_2b_1}\left(b_2\frac{d\Gamma^-_5}{dx}+b_1\frac{d\Gamma^-_6}{dx}\right)
\end{equation}
\begin{equation}
\label{eq39}
f^R_H\eta^I_P=\frac{1}{a_1b_2-a_2b_1}\left(a_2\frac{d\Gamma^-_5}{dx}+a_1\frac{d\Gamma^-_6}{dx}\right),
\end{equation}
from which we find
\vspace*{6mm}
\begin{equation}
\label{eq40}
\frac{f^I_H}{f^R_H}=\frac{b_2\,d\Gamma^-_5/dx+b_1\,d\Gamma^-_6/dx}{a_2d\Gamma^-_5/dx+a_1d\Gamma^-_6/dx}.
\vspace*{3mm}
\end{equation}
Since we are assuming that there is no
$Q^2,s_1$, or $s_2$ dependence in $f_H$, the right
hand side of Eq.~(\ref{eq38}) as well as of Eq.~(\ref{eq39}) must be
constant over the range of $x$. Therefore, by extracting the distributions
$d_x\Gamma^-_{5,6}$ from the data and obtaining the
quantities $a_{1,2},b_{1,2}$ numerically for each bin
in the $x$ range, the assumption
regarding the flatness of $f_H$ (as a function of
  $Q^2, s_1$ and $s_2$) can be tested. On the other hand, under the
assumption that $f_H$ has no strong phase,
the left hand side of Eq.~(\ref{eq40}) vanishes,
so that the significance of the
deviations from zero of the quantity appearing on the right hand side can be used to test this
assumption.\par Another possibility arises from the analysis of the
zero-crossing points for the various distributions. Under the
assumptions mentioned above, namely that $f^I_H=0$ and that its functional dependence on the kinematical variables is flat, the zero-crossing points for the
$\mathrm{CP}$-odd distributions are independent of the value of the NP
parameter $f_H\eta_P$. Thus, the
numerical prediction of these zero-crossing points and the comparison
with the distributions obtained from the data can also be used to test these two assumptions.\footnote{Here we are taking the parameters related to the resonance structure of the decay to be fixed to their input values. In fact, the position of the zero-crossing points depends not only on the two assumptions we are testing but also on these input values. In this sense, the analysis of the zero-crossing points could also be useful for studying these parameters.}
 In order to illustrate this, let us consider
the
\begin{figure}[H]
\centering
\includegraphics[width=0.56\textwidth,height=0.42\textwidth]
{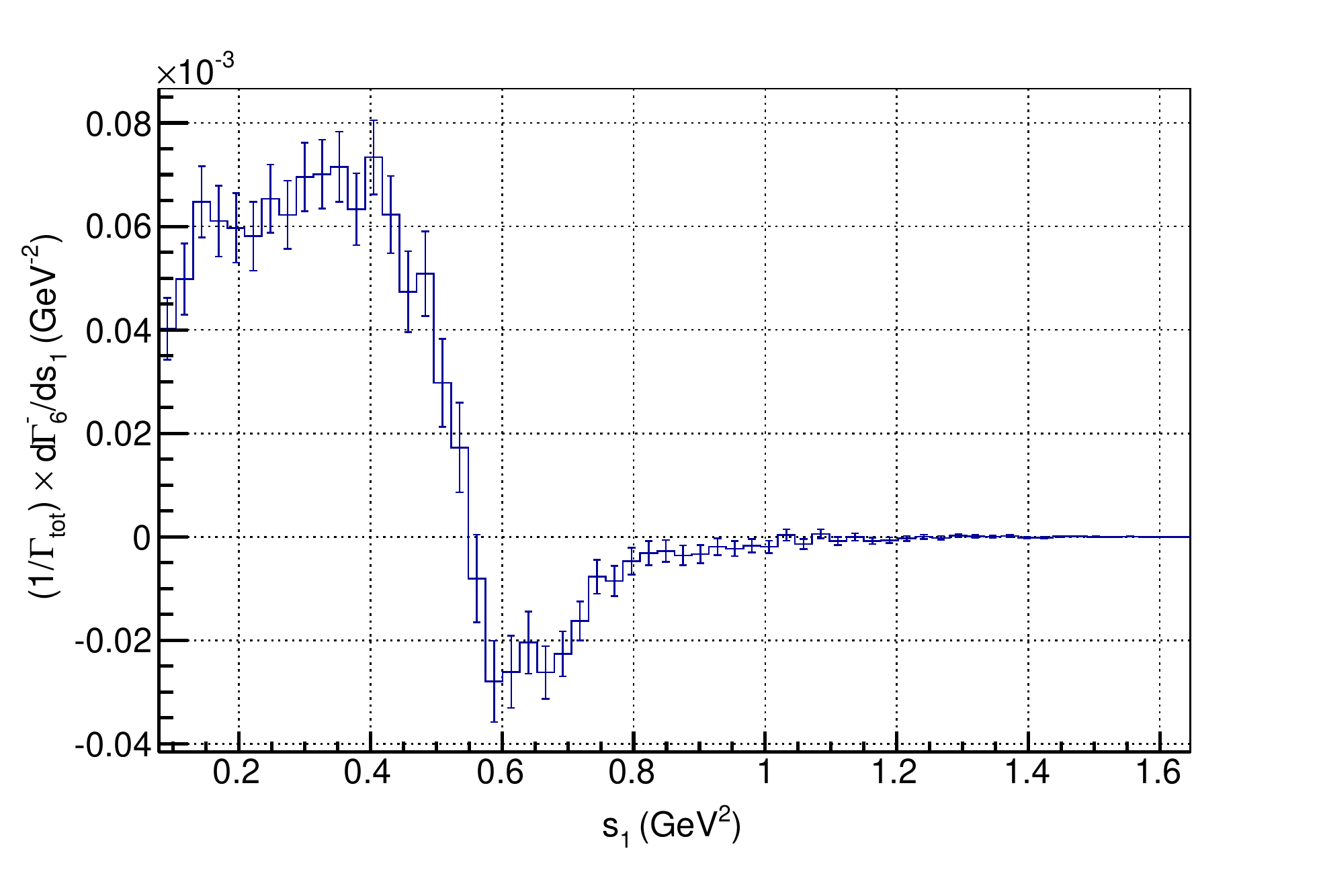}
\caption{Projection onto $s_1$ of the observable $6(-)$, obtained by using a set of $3\times 10^6$ simulated events. The zero-crossing point can be clearly extracted from the plot with an uncertainty given by the size of the bins ($s_1=0.56\pm 0.03\,\,\mathrm{GeV}^2$).}
\label{fig6oddsim}
\end{figure}
\noindent observable $6(-)$ (see Fig.~\ref{fig6odd}). Projecting this distribution separately onto $Q^2,s_1$ and $s_2$ and performing a numerical computation of the corresponding zero-crossing points yields the values $Q^2\sim 1.85609\,\,\mathrm{GeV}^2,\,s_1\sim 0.55633\,\,\mathrm{GeV}^2$ and $s_2\sim 0.85142\,\,\mathrm{GeV}^2$, respectively. On the other hand, 
analysis of the distributions associated with a set of $3\times 10^6$ events
yields the following values (see Fig.~\ref{fig6oddsim}) 
\begin{equation}
\vspace*{2mm}
\label{eq41}
Q^2=1.86\pm 0.04\,\,\mathrm{GeV}^2,\quad\,s_1=0.56\pm 0.03\,\,\mathrm{GeV}^2,\quad\,s_2=0.84\pm 0.04\,\,\mathrm{GeV}^2,
\end{equation}
which are in good agreement with the expected
values. 
Thus, with $3\times 10^6$ events, it appears that one could use the zero-crossing points of the $\mathrm{CP}$-odd distributions to test the assumptions regarding $f_H$ that were noted above.
With fewer than $3\times 10^6$ events, however, the zero-crossing point test would 
start to lose its effectiveness.
\section{Conclusions}
\label{conclusions}
In this paper we have proposed and tested various $\mathrm{CP}$-even
and $\mathrm{CP}$-odd observables for the decay $\tau\rightarrow K
\pi\pi\nu_{\tau}$ by adding the contribution of a NP charged scalar to
the corresponding amplitude within a model-independent approach.
The various observables that we have proposed are
  defined in Eq.~(\ref{eq14}) (see also Tables~\ref{tabla1} and
  \ref{tabla3}).  These observables are distributions that have been
  partially integrated over phase space, using weighting functions to
  pick out various terms from the original expression for the
  differential width (see Eq.~(\ref{eq5})).  The resulting
  distributions are functions of three invariant mass squared
  variables, $Q^2$, $s_1$ and $s_2$, and they depend on the NP
  contribution in different ways.  Throughout much of the text, we
  have denoted the various distributions by ``$i(\pm)$'' $(i = 1,
  \ldots, 9$), where the ``$\pm$'' designation refers to whether the
  distribution is even (``$+$'') or odd (``$-$'') under
  $\mathrm{CP}$. For the numerical analysis we have
used simulated events generated through our own event generator,
with the maximum number of simulated events being $3
\times 10^6$.\par

Among the various observables
that we have proposed, the $6(-)$
distribution is the most sensitive to the NP
contribution. On the one hand, for a
sizeable NP contribution ($|f_H\eta_P|\sim17.9$), we have found that
this observable is useful for testing the SM hypothesis, even for $1\times
10^5$ events. On the other hand, the results of the
fits show that this observable allows
one to recover the NP parameter with the highest
precision, with the uncertainties being
$\lesssim\!6\,\%$ and $\lesssim\!  3\,\%$ for $5\times 10^5$ and
$3\times 10^6$ simulated events, respectively. More
interestingly, the capability of the observable $6(-)$ to recover the
NP parameter is not spoiled when the size of the NP
contribution is reduced. \par

Regarding the $\mathrm{CP}$-even observables that we study in this paper, we have found that the
$5(+)$ distribution and the $s_{1,2}$ projections of
the $7(+)$ distribution
show the most promise for recovering the parameter
  $A$, which is related to the
weight of the resonant contributions.
Additionally, considering that the $8(+)$ and $9(+)$ distributions extracted from the set of $3\times 10^6$ simulated events are consistent with zero to within their statistical uncertainties, we have shown that the observable $5(+)$ is the most
suitable alternative for extracting information about the anomalous
Wess-Zumino term once the other parameters related to the various resonances have been
measured.\par
 The results involving the $\mathrm{CP}$-odd observables
have been derived under the assumptions that $f^I_H=0$ and that its functional dependence on the kinematical
  variables is flat. The same assumptions have been
  made for the $\mathrm{CP}$-even observables, but in that case,
 we have also assumed that $F_4=0$. The possibilities for testing some of these assumptions by using
the observables defined in this paper have been discussed in
Sec.\ref{sec7}.\par 

We have also studied the decay $\tauprocess$ within the context of the
A2HDM and have found that the
 observables that we have defined may
be used to test this model. In
particular, we have 
focused on the $s_2$ projection of the differential width $6(-)$, 
comparing the range allowed by the A2HDM to that predicted by
our simulation, adopting the NP parameter choice $|f_H\eta_P|=1.79$.  
Using a simulation with $10^6$ events, we have
found that the best fit value for $f_H\eta^I_P$ obtained from the distribution is in disagreement (by more than $2\sigma$) with the range predicted for the A2HDM.  
With the NP parameter choice
$|f_H\eta_P|=17.9$ and the same number of events, the
disagreement between the two scenarios is much greater and one
would be able to distinguish decisively between them.  \par
We note that a similar set of observables could be defined in order to
analyze other decay modes such as
$\tau^{-}\rightarrow
\pi^{-}\pi^{+}\pi^{-}\nu_{\tau},\,\tau^{-}\rightarrow
K^{-}K^{+}\pi^{-}\nu_{\tau}$ and $\tau^{-}\rightarrow
K^{-}K^{+}K^{-}\nu_{\tau}$, and their $\mathrm{CP}$-conjugated
decays. In fact, precise measurements of the branching
ratios for these decays have already been obtained at the B-factories (see
Refs.~\cite{Babar,Belle} for example). \par An experimental analysis of
the observables we have analyzed in this paper could be useful not only for extracting information about the resonance structure of the decay $\tauprocess$ but also for obtaining additional constraints on the NP pseudoscalar coupling. Moreover, with the higher
luminosity expected for the upcoming Super B-factories, the number of events
anticipated for the decay $\tauprocess$ would be enough to exploit the information provided by the proposed observables.

\bigskip
\noindent
{\bf Acknowledgments}
\noindent 
The authors wish to acknowledge helpful discussion and communication
with S. Banerjee, I. Nugent, M. Roney and G. Valencia.  They also wish
to thank C. Daudt, N. Lickey and N. White for technical assistance, and A. Pich for revising the manuscript.
This work has been partially supported by ANPCyT under grant
No. PICT-PRH 2009-0054 and by CONICET (NM, AS). The work of KK was
supported by the U.S.  National Science Foundation under Grant
PHY-1215785. KK also acknowledges sabbatical support from Taylor
University.

\appendix 
\section{Statistical Uncertainties}
\label{sec8}
In this appendix we summarize some results regarding statistical uncertainties
associated with the distributions considered in this work.

The estimator that we have used to extract the
projections onto $Q^2,s_1$, and $s_2$ of the weighted partial differential
widths from the simulated events is given by:
\begin{equation}
\label{Apeq1}
\frac{1}{\Gamma_{\mathrm{tot}}}\frac{\hat{d\Gamma}_i}{dx}(x_0)=\frac{N}{N_{\mathrm{ev}}}\frac{\bar{h}_i}{\Delta x}\mathcal{B}_{\tau\rightarrow K\pi\pi\nu_{\tau}} \,,
\end{equation}
where $\frac{\hat{d\Gamma}_i}{dx}(x_0)$ denotes the projection onto
$x\equiv Q^2,s_1,s_2$ of the $i$-th weighted partial width evaluated
at $x_0$, $N$ is the number of events within the bin $(x_0-\Delta
x/2,x_0+\Delta x/2)$, $\bar{h}_i$ is the sample mean of the angular
function $h_i(\gamma,\beta)$ (see Table \ref{tabla1}) in 
the bin and $N_{\mathrm{ev}}$ is the total number of
simulated events.  We note that the presence of the branching ratio ($\mathcal{B}_{\tau\rightarrow K\pi\pi\nu_{\tau}}$)
arises from the fact that we have normalized the observables to the
total decay width ($\Gamma_{\mathrm{tot}}$).\par
In order to estimate the statistical error
  associated with $d\Gamma_i/dx$, we use error propagation in Eq.~(\ref{Apeq1}), taking into account
  the standard deviations of the number of events in a given bin, $N$,
  and of the sample mean $\bar{h}_i$. The expression
  that we obtain for the $j$-th bin is given by:
\begin{equation}
\label{Apeq2}
\sigma_j =\frac{\mathcal{B}_{\tau\rightarrow K\pi\pi\nu_{\tau}}}{\Delta x}\frac{\sqrt{I_j}}{\sqrt{N_{\mathrm{ev}}}}(\sigma_{h_i}+\langle h_i\rangle\sqrt{1-I_j}),
\end{equation}
where $\sigma_{h_i}=\sqrt{\langle h^2_i \rangle - \langle h_i \rangle
  ^2}$ is the standard deviation of $h_i$ computed for the 
$j$-th bin, $I_j$ is the probability for a given
event to lie within that bin and $\langle
h^2_i\rangle$ and $\langle h_i\rangle$ denote the mean values of
$h^2_i$ and $h_i$, respectively, which are calculated, again, for the
$j$-th bin. In general, for all the
observables the dominant contribution arises from the standard
deviation of the angular function, $\sigma_{h_i}$, while the second
term in Eq.~(\ref{Apeq2}) is negligible. The unique exception is the
observable with $i=1$, for which $\sigma_{h_1}=0$ (due to the fact
that $h_1(\alpha,\beta)=1$ -- see Table \ref{tabla1}),
so that the second term is the dominant
one. Actually, this second term computed for the observable
$d\Gamma_1/dx$ turns out to be comparable to the first contribution
obtained for any of the remaining observables ($d\Gamma_i/dx,\,
i=2,\ldots, 9$). Therefore, the statistical uncertainties $\sigma_j$ are of
the same order of magnitude for all of the weighted
partial widths ($i=1,\ldots, 9$). Of course, the order of magnitude of
the uncertainty in Eq.~(\ref{Apeq2}) changes from one bin to another
and from one projection to another ($x=\,Q^2,s_1$ or $s_2$).
\section*{\refname}
\let\bibsection\relax

\setlength{\bibsep}{10pt}
\bibliography{Paperbiblio}
\bibliographystyle{apsrev4-1}

\end{document}